\documentclass{aa}  

\usepackage{natbib}
\bibpunct{(}{)}{;}{a}{}{,}
\usepackage[colorlinks=true,linkcolor=blue,citecolor=blue,filecolor=magenta,urlcolor=cyan]{hyperref}
\usepackage{graphicx}
\usepackage{txfonts}
\usepackage{enumitem}
\usepackage{tikz}
\usetikzlibrary{spy}

\definecolor{Sienna}{rgb}{0.53, 0.18, 0.09}

\graphicspath{{figures/}}
\def\msun{\!{\rm M}_{\sun}}

\def\lsim{\mathrel{\rlap{\lower 3pt \hbox{$\sim$}} \raise 2.0pt \hbox{$<$}}}
\def\gsim{\mathrel{\rlap{\lower 3pt \hbox{$\sim$}} \raise 2.0pt \hbox{$>$}}}

\defcitealias{Sassano_et_al_2023}{S23}

\usepackage{orcidlink}

\begin{document} 

\title{Super-Eddington accretion in protogalactic cores}

\author{Tommaso~Zana\inst{1,2,3}\fnmsep\thanks{Corresponding author; tommaso.zana@uniroma1.it}\orcidlink{0000-0003-4244-8527} \and
    Pedro~R.~Capelo\inst{4}\orcidlink{0000-0002-1786-963X} \and
    Mairo~Boresta\inst{1}\orcidlink{0009-0007-7581-1747} \and
    Raffaella~Schneider\inst{1,2,3,5}\orcidlink{0000-0001-9317-2888} \and
    Alessandro~Lupi\inst{6,7,8}\orcidlink{0000-0001-6106-7821} \and
    Alessandro~Trinca\inst{6,2,3}\orcidlink{0000-0002-1899-4360} \and
    Lucio~Mayer\inst{4}\orcidlink{0000-0002-7078-2074} \and
    Rosa~Valiante\inst{2}\orcidlink{0000-0003-3050-1765} \and
    Luca~Graziani\inst{1,3}\orcidlink{0000-0002-9231-1505}
}

\institute{
    Dipartimento di Fisica, Sapienza, Universit\`a di Roma, Piazzale Aldo Moro 5, IT-00185 Roma, Italy
    \and
    INAF, Osservatorio Astronomico di Roma, Via di Frascati 33, IT-00078 Monte Porzio Catone, Italy
    \and
    INFN, Sezione di Roma I, Piazzale Aldo Moro 2, IT-00185 Roma, Italy
    \and
    Department of Astrophysics, University of Zurich, Winterthurerstrasse 190, CH-8057 Zürich, Switzerland
    \and
    Sapienza School for Advanced Studies, Viale Regina Elena 291, IT-00161 Roma, Italy
    \and
    Como Lake Center for Astrophysics, DiSAT, Universit\`a degli Studi dell'Insubria, Via Valleggio 11, IT-22100 Como, Italy
    \and
    INFN, Sezione di Milano-Bicocca, Piazza della Scienza 3, IT-20126 Milano, Italy
    \and
    INAF, Osservatorio di Astrofisica e Scienza dello Spazio di Bologna, Via Gobetti 93/3, IT-40129 Bologna, Italy
}
\date{Received 27 August 2025 / Accepted 10 February 2026}

\titlerunning{Super-Eddington accretion in protogalactic cores}
\authorrunning{Zana T. et al.}

\abstract 
   {The presence of massive black holes (BHs) exceeding $10^9~\msun$ already at redshift $z > 6$ challenges standard models of BH growth. Super-Eddington (SE) accretion has emerged as a promising mechanism to solve this issue, yet its impact on early BH evolution in tailored numerical experiments remains largely unexplored. In this work, we investigate the growth of BH seeds embedded in a gas-rich, metal-poor protogalaxy at $z \sim 15$ using a suite of high-resolution hydrodynamical simulations that implement a slim-disc-based SE accretion model. We explored a broad parameter space, varying the initial BH mass, feedback efficiency, and spin. We find that SE accretion enables rapid growth in all cases, allowing BHs to accrete up to $10^5~\msun$ within a few $10^3$--$10^4$ years, independent of seed properties. Feedback regulates this process, both by depleting central gas and altering BH dynamics via star formation-driven potential fluctuations, yet even the strongest feedback regimes permit significantly greater growth than the Eddington-limited case. Growth stalls after less than $\sim$1 Myr due to local gas exhaustion, as no large-scale inflows are present in the adopted numerical setup. Our results show that SE accretion naturally leads to BHs that are overmassive relative to their host galaxy stellar content, consistent with JWST observations. We conclude that short low-duty-cycle SE episodes represent a viable pathway for assembling the most massive BHs observed at early cosmic times, even when starting from light seeds.}


\maketitle


\section{Introduction}
Massive black holes (BHs) are very common throughout the Universe. They reside at the centres of nearly all galaxies and are observable up to redshift $z > 7$. Their masses can reach $10^9$--$10^{10}~\msun$ \citep[][]{Fan_et_al_2001, Fan_et_al_2006, Mortlock_et_al_2011, Banados_et_al_2018, Yang_et_al_2020, Wang_et_al_2021}. These massive BHs typically shine as active galactic nuclei (AGNs), powered by gas accretion, which redistributes energy and momentum into the surrounding environment with a certain radiative efficiency, $\epsilon_{\rm r} \equiv L/\dot{M}_{\rm acc}c^2$, where $\dot{M}_{\rm acc}$ is the mass accreted per unit time onto the BH, $c$ is the speed of light in vacuum, and $L$ is the bolometric luminosity of the AGN. 

Observations of luminous AGNs at such high redshift suggest that BHs with mass $M_{\rm BH} \sim 10^9~\msun$ were already in place when the Universe was less than 800 million years old. Given the evolution of the AGN luminosity function and So\l{}tan's argument \citep[][]{Soltan_1982}, these BHs most likely grew from less massive seeds \citep[e.g.][]{Inayoshi_et_al_2020}, with the majority of their mass acquired through gas accretion, as the coalescence of BHs is less efficient in the long run \citep[e.g.][]{Kormendy_Ho_2013, Kulier_et_al_2015}.
Under the previous hypothesis, the BH growth rate can be described by
\begin{equation}
    \frac{{\rm d}M_{\rm BH}}{{\rm d}t}=(1-\epsilon_{\rm r})\dot{M}_{\rm acc}.
    \label{eq:BHgrowth}
\end{equation}

Assuming the maximum standard accretion rate defined by the \citet{Eddington_1916} limit given by
\begin{equation}
   \dot{M}_{\rm Edd} = \frac{4\pi G M_{\rm BH} \mu_{\rm e} m_{\rm p}}{\sigma_{\rm T} \epsilon_{\rm r} c},
   \label{eq:EddLim}
\end{equation}
where $G$ is the gravitational constant, $\mu_{\rm e}$ is the mean molecular weight per electron, $m_{\rm p}$ is the mass of the proton, and $\sigma_{\rm T}$ is the \citet{Thomson_1906} scattering cross section, Eq.~\eqref{eq:BHgrowth} leads to an exponential growth law with a characteristic timescale given by the \citet{Salpeter_1964} time:
\begin{equation}
    \tau_{\rm acc}= \frac{\epsilon_{\rm r}}{1-\epsilon_{\rm r}}\frac{\sigma_{\rm T}c}{4\pi G \mu_{\rm e} m_{\rm p}}.
    \label{eq:tau}
\end{equation}
Here, the standard radiative efficiency for a sub-Eddington thin disc model depends on the BH spin. For a non-rotating \citet{Schwarzschild_1916} BH, $\epsilon_{\rm r}$ is approximately $0.06$, while for rapidly spinning \citet{Kerr_1963} BHs, $\epsilon_{\rm r}$ can reach values as high as $\sim$20\% for magnetised disc accretions \citep[][]{Krolik_et_al_2005} and even $\sim$30\% for unmagnetised thin discs \citep[][]{Thorne_1974}.
If we consider the accreted gas to be primordial and fully ionized with $\mu_{\rm e} = 1.15$ and the typically adopted $\epsilon_{\rm r} = 0.1$ \citep[][]{Soltan_1982}, representing a moderately rotating BH, then the accretion timescale is $\tau_{\rm acc}=4.35 \times 10^{7}$~yr.
For example, a BH with an initial mass of $500~\msun$ would require approximately $15\,\tau_{\rm acc}$, assuming a constant specific accretion rate at the Eddington limit, to form a massive BH of $10^{9}~\msun$. This timescale is longer than the cosmic time interval between $z \sim 24$ and $z \sim 7$ in a $\Lambda$ cold dark matter (DM) cosmology with $\Omega_{\rm M}= 0.3089$, $\Omega_{\Lambda} = 0.691$, and $h = 0.677$ \citep[][]{Planck_2016}.
Interestingly, even in the case of a massive initial seed of $10^{5}~\msun$ from the direct collapse scenario \citep[e.g.][]{Begelman_et_al_2006, Patrick_et_al_2023}, the constant specific accretion rate at $\dot{M}_{\rm Edd}$ would take about $9\,\tau_{\rm acc}$, i.e. longer than the time between $z \sim 12$ and $z \sim 7$.
This indicates that even in the extreme case of a large initial mass, a BH would still need to accrete continuously at the maximum allowed rate for nearly 400 million years -- a scenario that is highly unlikely, particularly due to stellar feedback \citep[e.g.][]{Dubois_et_al_2015, Habouzit_et_al_2017, Angel-Alcazar_et_al_2017}.
To further complicate matters, of all the models for the formation of BH seeds, mainstream\footnote{Alternative direct collapse scenarios exist that do not rely on specific environmental conditions or metallicity, and they are instead triggered by gas inflows in high-redshift galaxy mergers, which can generate supermassive stars or clouds unstable to the relativistic radial instability \citep[][]{Mayer_et_al_2010, Zwick_et_al_2023, Mayer_et_al_2024}. However, the required high inflow rates are expected to occur only in the most massive galaxies, consistent with the population of bright high-redshift quasars but likely insufficient to explain the abundant population of lower-luminosity AGNs recently discovered by JWST at even higher redshifts. Interestingly, \citet{van_Dokkum_et_al_2025} have recently reported a system that appears to be a compelling example of such a process.} direct collapse shows some problems, as it requires specific and stringent conditions to avoid fragmentation \citep[e.g.][but see also \citealp{Chon_Omukai_2025}]{Omukai_2001,Wise_et_al_2019}.
Indeed, PopIII remnants ($\sim$$10^{2}~\msun$ BHs, formed at $z\gtrsim20$) appear to be the most natural sources of BH seeds, given their widespread presence in early DM mini-haloes, while the medium-weight scenario ($\sim$$10^3$--$10^4~\msun$ BHs formed from stellar collisions within a nuclear star cluster) offers an appealing alternative, as it does not rely on finely tuned cosmological conditions, and many cluster are found to be common even at very high redshift (see e.g. \citealt{Adamo_et_al_2024} and \citealt{Volonteri_2010} for a review).
Although both of these less massive initial seeds are more likely, they still challenge the growth scenario, as they would require accreting more mass compared to the direct collapse case.

A promising solution would be to allow BHs to grow episodically at a rate above the Eddington limit. Super-Eddington (SE) accretion has been directly observed in low-redshift phenomena such as tidal disruption events \citep[][]{Rees_1988, Lin_et_al_2017}, ultraluminous X-ray sources \citep[][]{Begelman_2002, Begelman_et_al_2006, Bachetti_et_al_2014}, and AGN \citep[][]{Du_et_al_2014, Du_et_al_2018, Tortosa_et_al_2023}. More recently, SE models have gained significant recognition as a possible explanation for the rapid early growth of BHs \citep[e.g.][]{Schneider_et_al_2023, Lupi_et_al_2024b, Trinca_et_al_2024, Madau_Haardt_2024, Inayoshi_Maiolino_2025}, particularly in light of the new wealth of AGNs observed with the James Webb Space Telescope \citep[JWST; see e.g.][]{Yang_et_al_2023, Ono_et_al_2023, Ubler_et_al_2023, Harikane_et_al_2023, Maiolino_et_al_2024_GNz11, Juodzbalis_et_al_2024}. The same observations suggest that a substantial fraction of the AGNs detected by JWST grew at a faster rate compared to the stellar component of their host galaxies, resulting in BHs that appear overmassive relative to their host stellar component \citep[][]{Bogdan_et_al_2024, Furtak_et_al_2024, Ubler_et_al_2023, Maiolino_et_al_2024, Harikane_et_al_2023, Juodzbalis_et_al_2026} when compared to what is inferred from the local $M_{\rm BH}$--$M_*$ relation \citep[][]{Kormendy_Ho_2013, Reines_Volonteri_2015}.

Super-Eddington accretion can naturally occur under realistic astrophysical conditions, for instance when the spherical symmetry required by the Eddington limit is broken, and photon propagation and gas inflows occur in different directions \citep[e.g.][]{Krumholz_et_al_2009}. 
In particular, radiation can be emitted highly anisotropically due to intrinsic production mechanisms or opacity inhomogeneities within the surrounding medium \citep[][]{Begelman_2002, Begelman_et_al_2006}.
Alternatively, when the density of the accreting gas is sufficiently high, advection dominates over radiative diffusion, resulting in photon trapping. In this case, photons are effectively accreted inwards along with the gas, as described by the commonly adopted slim-disc model \citep[e.g.][]{Abramowicz_et_al_1988, Kato_et_al_2004, Sadowski_2009} and its subsequent refinements \citep[e.g.][]{Jiang_et_al_2014, Jiang_et_al_2019}.
Photon trapping leads macroscopically to a reduction in radiative efficiency.\footnote{We note that alternative hypotheses exist, suggesting that SE accretion can also occur while maintaining high radiative efficiency \citep[e.g.][]{Socrates_Davis_2006}.}
From Eq.~\eqref{eq:tau}, one can see that $\tau_{\rm acc} \propto \epsilon_{\rm r}/(1 - \epsilon_{\rm r})$. Thus, any process capable of significantly reducing the radiative efficiency could dramatically enhance the accretion rate. 

Apart from hyper-resolved magneto-hydrodynamical simulations that focus on small-scale regions around BHs and incorporate general relativity \citep[e.g.][]{Sadowski_et_al_2016, Dai_et_al_2018, Curd_et_al_2019}, which provide crucial insights but remain computationally prohibitive for simulating even just the entire accretion disc, some recent numerical models have started to adopt specialised sub-grid prescriptions to effectively describe SE accretion within complex astrophysical environments, from galactic cores to cosmological simulations. 
However, these models are often overly simplistic, merely extending standard accretion and feedback prescriptions beyond the Eddington limit without accounting for the complex physical processes involved \citep[e.g.][]{Beckmann_et_al_2019, Ni_et_al_2022, Zhu_et_al_2022, Bhowmick_et_al_2022, Gordon_et_al_2025}.
Among the more physically motivated models are state-of-the-art schemes that include prescriptions for the self-consistent launching of winds and jets in large environments, such as those presented by \citet{Lupi_et_al_2024a} and \citet{Husko_et_al_2025} in cosmological contexts. However, these approaches necessarily sacrifice resolution, limiting their ability to accurately track the accretion process in the immediate vicinity of the BH.
On the other hand, less detailed prescriptions are typically adopted in higher-resolution numerical studies that focus on much smaller spatial regions, enabling a more accurate treatment of gas dynamics in the nuclear region of the host galaxy \citep[e.g.][]{Lupi_et_al_2016, Sassano_et_al_2023, Massonneau_et_al_2023, Toyouchi_et_al_2024, Mehta_et_al_2024, Kao_et_al_2025}.

Notably, some semi-analytical models have recently incorporated SE accretion in their studies of galaxy evolution, particularly at high redshift \citep[e.g.][]{Trinca_et_al_2022, Schneider_et_al_2023, Izquierdo-Villalba_et_al_2024, Trinca_et_al_2024}.
In particular, \citet{Trinca_et_al_2024} successfully reproduced both the observed BH masses and the deviation from the local $M_{\rm BH}$--$M_*$ relation observed at high redshift by considering sporadic episodes of SE accretion starting from different initial seed masses.

Here, we investigate in detail the immediate impact of SE accretion on the early growth of different BH seeds located at the centre of a metal-poor gas-rich protogalactic system, representative of a typical high-redshift environment host.
By exploring a wide parameter space, we assess the role of the initial BH seed mass, feedback intensity, and BH spin in modulating the accretion efficiency.
Our goal is to provide numerical support for recent findings from semi-analytical models and observations and to evaluate the plausibility of short bursts of SE accretion as a viable growth channel.

A detailed description of our simulation suite is provided in Sec.~\ref{sec:simulations}, whereas Sec.~\ref{sec:results} presents the results. In Sec.~\ref{sec:discussion}, we summarise and discuss the implications of our findings. We draw our conclusions in Sec.~\ref{sec:conclusion}.

\section{The simulation suite}
\label{sec:simulations}

For the purpose of our investigation, we construct an idealised environment and run a suite of high-resolution smoothed-particle hydrodynamics (SPH) simulations, exploring the dependence of BH growth on the initial seed mass, feedback strength, and BH spin.

\subsection{The code}
\label{subsec:code}

All simulations presented in this work were run with an updated version of \textsc{Gasoline2} \citep[][]{Wadsley_et_al_2017}, a parallel $N$-body SPH code based on the $N$-body code \textsc{pkdgrav} \citep[][]{Stadel_2001} and descendant of the \textsc{Gasoline} code \citep[][]{Wadsley_et_al_2004}. \textsc{Gasoline2} presents several improvements over \textsc{Gasoline}, including explicit turbulent diffusion terms \citep[][]{Wadsley_et_al_2008, Shen_et_al_2010}, time-dependent local viscosity limiters \citep[][]{Morris_Monaghan_1997, Cullen_Dehnen_2010}, upgraded kernels \citep[][]{Wendland_1995, Dehnen_Aly_2012, Keller_et_al_2014}, a geometric density average force expression \citep[][]{Monaghan_1992, Ritchie_Thomas_2001, Keller_et_al_2014}, and gradient-based shock detection \citep[][]{Wadsley_et_al_2017}.

The code computes non-equilibrium abundances and cooling for primordial species, assuming gas self-shielding and a radiation background \citep[][]{Pontzen_et_al_2008, Haardt_Madau_2012}, and photo-ionisation equilibrium cooling from metal fine structure lines \citep[][]{Shen_et_al_2010, Shen_et_al_2013}, using rates from \textsc{Cloudy} \citep[][]{Ferland_et_al_1998, Ferland_et_al_2013} and assuming no self-shielding \citep[for a discussion, see][]{Capelo_et_al_2018}.\footnote{We note that, together with the metal cooling described in \citet{Shen_et_al_2010,Shen_et_al_2013}, we employed the low-temperature metal cooling by \citet{Mashchenko_et_al_2008}, itself a fit for $T < 10^4$~K of the cooling function of \citeauthor{Ricotti_et_al_1997} (\citeyear{Ricotti_et_al_1997}; see also \citealt{Bromm_et_al_2001}). We assessed the importance of this additional metal cooling at the simulated redshift ($z = 15$) and verified that the cooling function from the equilibrium tables is always much larger than that from the low-temperature fit, except at high gas densities. Specifically, the cooling functions are comparable only at densities $\gtrsim$100$~m_{\rm p}$~cm$^{-3}$, hence well above the SF density threshold (see text).}
Star formation (SF) follows the recipe of \citet{Stinson_et_al_2006}, wherein star particles \citep[proxies of stellar populations with a given initial mass function;][]{Kroupa_2001} can be stochastically generated in a sufficiently cold ($< 1.5 \times 10^4$~K in this work) and dense ($> 5 \,m_{\rm p}$~cm$^{-3}$) gas, according to a probability function consistent with the \citet{Schmidt_1959} law. Stellar feedback, in the form of supernova explosions and stellar winds \citep[][]{Stinson_et_al_2006}, is also implemented, although it is not relevant for this work, as the simulation timescale is shorter than that of feedback; moreover, even if stellar winds were present, their impact would be further reduced by the very low metallicity of the environment.

In the code of \citep[][]{Bellovary_et_al_2013}, the BHs are implemented as sink particles that accrete the surrounding gas according to the Bondi-Hoyle-Lyttleton \citep[hereafter BHL;][]{Hoyle_Lyttleton_1939, Bondi_Hoyle_1944, Bondi_1952} accretion rate, defined as
\begin{equation}
    \dot{M}_{\rm BHL} = \frac{4 \pi \alpha G^2 M_{\rm BH}^2 \rho_{\rm gas}}{(v^2 + c_{\rm s}^2)^{3/2}}~,
    \label{eq:BHL}
\end{equation}
where $v$ is the local velocity of the gas with respect to the BH, $c_{\rm s}$ is the local speed of sound, and $\alpha$ is the so-called boost factor, usually adopted in low-resolution simulations, which we set equal to 1 \citep[as in other high-resolution studies using \textsc{Gasoline}/\textsc{Gasoline2}, e.g.][]{Capelo_et_al_2015, SouzaLima_et_al_2017}. More specifically, the BHL accretion rate is computed for each of the $N_{\rm smooth}$ neighbour gas particles in the BH kernel (64 in this work), then the mass is removed from each gas particle according to its contribution, and the combined mass is added to the BH according to Eq.~\eqref{eq:BHgrowth}.

In the standard version of the code, the BH accretion rate is capped at the Eddington mass accretion rate, given by Eq.~\eqref{eq:EddLim}, where $\epsilon_{\rm r}$ has a fixed value, set in the parameter file, usually equal to 0.1, and $\mu_{\rm e}$ is set equal to unity for simplicity.

Part of the radiated luminosity, $L = \epsilon_{\rm r}\dot{M}_{\rm BHL}c^2$, is coupled isotropically\footnote{Although feedback from accreting BHs is expected to be highly anisotropic on very small scales, the energy and momentum output is generally thought to become more isotropized after thermalisation. Moreover, the effective angular distribution of the outflow is ultimately shaped by the ambient gas density, as the outflow propagates preferentially along paths of least resistance \citep[see for details, e.g.][]{Mukherjee_et_al_2018, Costa_et_al_2020, Tanner_Weaver_2022}.}
 to the surrounding gas. In \textsc{Gasoline2}, this so-called BH feedback is thermal, i.e. a small fraction $\epsilon_{\rm c}$ of $L$ is converted into thermal energy per unit time and coupled to the $N_{\rm smooth}$ neighbour gas particles in the BH kernel \citep[][]{Bellovary_et_al_2013}.

In \citeauthor{Sassano_et_al_2023} (\citeyear{Sassano_et_al_2023}; hereafter \citetalias{Sassano_et_al_2023}), the code has been modified in order to study SE accretion, removing the Eddington cap and adopting a variable radiative efficiency consistent with the slim disc model \citep[][]{Abramowicz_et_al_1988}. More specifically, they adopted the results by \citet{Madau_et_al_2014}, who provide an analytical fit of the numerical solutions of the relativistic slim accretion disc equations by \citet{Sadowski_2009}:
\begin{equation}
    \epsilon_{\rm r} = \frac{A(a)}{16} \left[ \frac{0.985}{1+B(a)\frac{\dot{M}_{\rm acc}c^2}{16 L_{\rm Edd}}} + \frac{0.015}{1+C(a) \frac{\dot{M}_{\rm acc}c^2}{16 L_{\rm Edd}}} \right],
    \label{eq:radiative_efficiency}
\end{equation}
where A, B, and C are functions of the BH spin $a$, defined as
\begin{equation}
    a \equiv cJ_{\rm BH}/(GM_{\rm BH}^2),
    \label{eq:BHspin}
\end{equation}
with $J_{\rm BH}$ being the BH angular momentum. The functions A, B, and C are defined as
\begin{equation}
    \begin{aligned}
        A(a) &= (0.9663 - 0.9292\,a)^{-0.5639},\\
        B(a) &= (4.627  - 4.445\,a)^{-0.5524},\\
        C(a) &= (827.3  - 718.1\,a)^{-0.7060},
    \end{aligned}
    \label{eq:radiative_efficiency_coefficients}
\end{equation}
respectively.\footnote{In \citetalias{Sassano_et_al_2023}, the authors also explored the radiative efficiency parametrization proposed by \citet{Jiang_et_al_2014}.}

In this work, we adopted a BH spin that is held constant for the entire simulation. While this choice is not fully physical, simulations that self-consistently evolve the BH angular momentum over a timescale comparable to ours show that the spin parameter changes only modestly -- even in the presence of substantial gas inflow and SE accretion \citep[see e.g. the gas-disc misaligned simulations in][]{Kao_et_al_2025}.
However, we explore two limiting cases -- a rapidly rotating BH ($a = 0.99$) and a non-rotating one ($a = 0$) -- to bracket the plausible range of radiative efficiencies and quantify the impact of the spin on our results (see Sec.~\ref{subsec:ic}).

For this work, we have substantially modified and improved the version of \textsc{Gasoline2} adopted in \citetalias{Sassano_et_al_2023}, which contained some inconsistencies in the BH prescription. All details of these
improvements are provided in Appendix~\ref{sec:code_improvements}.

\subsection{The initial conditions}
\label{subsec:ic}

We seeded BHs at the initial time $t_{\rm i} = 0$ with a mass $M_{\rm BH}(t_{\rm i})$ equal to $5 \times 10^2$, $10^3$, $5 \times 10^3$, $10^4$, and $10^5~\msun$ at the centre of a DM halo of mass $M_{\rm DM} = 2.5 \times 10^7~\msun$, representing the inner region (50\% of the total mass) of a typical galactic host at $z \sim 15$, analogously to what has been done in \citetalias{Sassano_et_al_2023}. We sampled the DM component with $10^6$ particles, distributed following a \citet{Plummer_1911} sphere, with a scale radius of 18~pc and a maximum radius of approximately 184~pc. The DM halo is set in preliminary equilibrium with a fully gaseous disc of mass $5 \times 10^6~\msun$, sampled with $2 \times 10^5$ particles and distributed according to a \citet{Mestel_1963} profile, with a scale height-to-radius ratio of $H/R = 0.05$ and a maximum outer radius of 55~pc.
At this point, the initial metallicity is set to $10^{-3}~Z_{\sun}$, and the initial temperature to 2500~K, ensuring that the disc is Toomre stable \citep[][]{Toomre_1964} and thus preventing early SF.
Subsequently, we perform an adiabatic relaxation of all the systems, each in equilibrium with a different BH seed, by evolving it without radiative cooling or SF for approximately 10~Myr -- equivalent to about seven orbital times at $R=10$~pc -- allowing the system to reach full dynamical equilibrium between the BH seed and its surrounding environment.\footnote{This final step, which was missing in the work by \citetalias{Sassano_et_al_2023}, plays a significant role in the initial phases of BH accretion.}
In Fig.~\ref{fig:map_ic}, we show the surface density maps of the unperturbed relaxed gas disc.

We adopted a mass resolution of $25~\msun$ and a gravitational softening length of 0.18~pc \citep[i.e. an equivalent Plummer softening length of 0.13~pc; see][]{Kim_et_al_2016}, which -- under our typical gas temperature conditions -- is comparable to the \citet{Bondi_1952} radius of the central BH for the majority of the time. This setup ensures that we can accurately resolve the dynamics in the vicinity of the BH.

\begin{figure*}
    \centering
    \includegraphics[width=1\linewidth]{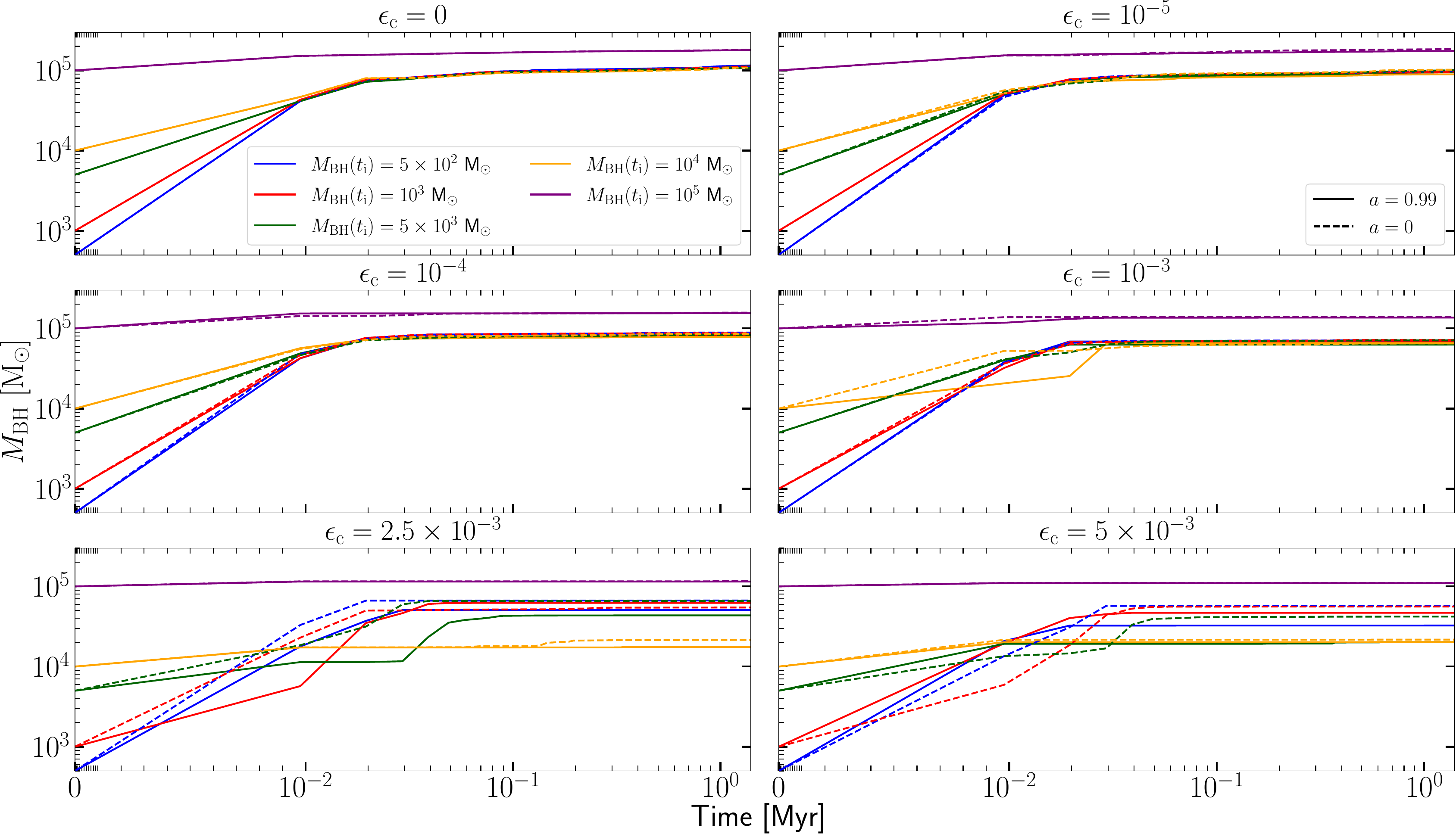}
    \caption{Evolution of the BH mass for all the tested BH seeds, grouped by $\epsilon_{\rm c}$. Different colours indicate the initial BH mass, whereas the line style denotes the BH spin: solid lines for $a = 0.99$ and dashed lines for $a = 0$. The $x$-axis is shown using a symmetric logarithmic scale with a linear threshold of $10^{-2}$ to allow visibility near $t = 0$ while preserving a logarithmic behaviour at larger values.}
    \label{fig:bhmass_evol_eps}
\end{figure*}

We assumed that no SF occurs prior to the seeding of the BH. We set the temperature cooling floor to $T_{\rm min} = 44$~K, corresponding to the temperature of the cosmic microwave background at the simulated redshift ($z = 15$). This choice ensures that the \citet{Jeans_1902} mass is always resolved by at least $N_{\rm smooth}$ gas particles for all densities below $\sim$$3 \times 10^3 \,m_{\rm p}$~cm$^{-3}$.

To explore the impact of BH feedback, we perform a set of tests varying the coupling efficiency $\epsilon_{\rm c}$, adopting values of 0, $10^{-5}$, $10^{-4}$, $10^{-3}$, $2.5 \times 10^{-3}$, and $5 \times 10^{-3}$. 
This grid was chosen ($i$) to remain close to the values adopted by \citetalias{Sassano_et_al_2023}, whose setup is similar to ours and who found that BH growth is maximised for $\epsilon_{\rm c}=10^{-5}$, and ($ii$) to enable a direct comparison with the SE study of \citet{Trinca_et_al_2024} based on the semi-analytic CAT model,\footnote{We are aware that the coupling efficiency in our high-resolution simulations cannot be directly compared to its counterpart in CAT, mainly because of different sub-grid implementations and, in particular, the characteristic timescales involved. Nevertheless, using the same parameter allowed us to directly measure how BH evolution responds in the two different approaches.} which assumes a fixed $\epsilon_{\rm c} = 2.5 \times 10^{-3}$. Finally, each simulation was run twice, once with $a = 0.99$ and once with $a = 0$, to gauge the impact of the BH spin on the accretion.

Overall, we ran simulations that cover every permutation of the quantities $M_{\rm BH} = 5 \times 10^2$, $10^3$, $5 \times 10^3$, $10^4$, $10^5$~M$_{\sun}$, $\epsilon_{\rm c} = 0$, $10^{-5}$, $10^{-4}$, $10^{-3}$, $2.5 \times 10^{-3}$, and $5 \times 10^{-3}$, and $a = 0$ and 0.99. On top of these 60 simulations, we also re-ran a subset of them in the Eddington limited (EL) case, disabling SF, and raising the cooling temperature floor (see Sec.~\ref{subsec:stalled_growth} for more details).

\section{Results}
\label{sec:results}

\subsection{Growth of massive black holes}

In Fig.~\ref{fig:bhmass_evol_eps}, we report the evolution of the accreted mass for our entire suite of runs, from $t_{\rm i} = 0$ to the end of the simulations $t_{\rm f} \simeq 1.5$~Myr, grouped by coupling efficiency $\epsilon_{\rm c}$, and for the two spin limit cases, $a = 0.99$ (solid lines) and $a = 0$ (dashed lines).
All runs show that a rapid and significant growth of $M_{\rm BH}$, lasting a few kiloyears, is always followed by a period of negligible (but non-zero) mass growth, independent of the initial mass $M_{\rm BH}(t_{\rm i})$, the feedback intensity, or the BH spin.
As a result, BHs quickly reach an almost constant value allowed by the immediate environment, determined here by the initial conditions, namely the total gas mass and its density distribution.

Feedback has a significant impact on the accretion process. It is evident that, as the feedback intensity increases, the evolutionary paths of BHs with different seed masses begin to diverge more markedly (though this divergence is not linear, and the final masses do not appear to retain a clear dependence on the initial seed mass). In particular, for feedback efficiencies up to $\epsilon_{\rm c} = 10^{-4}$, in the low-feedback regime, all BHs tend to converge towards similar final masses, quickly losing memory of their initial value.
An exception to this behaviour is the case with $M_{\rm BH}(t_{\rm i})=10^5~\msun$, which shows no substantial growth relative to its initial mass. This is due to the limited availability of gas, which not only prevents the BH from significantly increasing its already large initial mass, but also restricts less massive seeds from reaching similar values.

In higher-feedback regimes -- specifically when $\epsilon_{\rm c} > 10^{-4}$, and increasingly so with rising intensity -- we observe that some seeds exhibit reduced growth, generally, though not always, those with the highest initial masses. 
Once again, the seed with $M_{\rm BH}(t_{\rm i}) = 10^5~\msun$ follows a somewhat distinct evolutionary path, as it is too massive for the host system considered here and was included in the analysis as a limiting case.
This apparent lesser growth of the more massive seeds is mainly due to two reasons.
($i$) At the onset of the simulation, when feedback is particularly strong, because of the high BH activity, a more massive BH produces more powerful outflows. These may succeed in expelling the majority of the available gas required for sustained accretion. This is especially true for the limiting case with $M_{\rm BH}(t_{\rm i}) = 10^5~\msun$, wherein nearly all accretion occurs within the first few thousand years. After this initial phase, no further significant accretion episodes take place, as a large low-density cavity is rapidly formed around the BH.
($ii$) As the coupling efficiency increases, the BH evolution becomes more chaotic. This is because strong feedback increasingly affects SF in the vicinity of the compact object, altering both the mass distribution and the orbital velocities of the newly formed stellar clusters. These factors, in turn, influence the BH dynamical behaviour, ranging from pinning the BH near the centre of the gravitational potential (as is the case, for instance, of runs without feedback), to scattering it a few parsecs away from the centre (in the maximum feedback case). This behaviour directly affects the availability of gas for BH accretion. However, it is only a second-order effect, as it mainly influences later stages of growth rather than the initial accretion episode.

A final remark concerns the effect of BH spin. Differences are generally minimal, however -- when present -- we observe that the cumulative impact of a high spin is to reduce the final mass of the BH, owing to the increase in the resulting radiative efficiency (see Eqs~\ref{eq:radiative_efficiency} and~\ref{eq:radiative_efficiency_coefficients} or, more clearly, fig.~1 from \citealt{Madau_et_al_2014}).
As shown in Fig.~\ref{fig:bhmass_evol_eps}, even the magnitude of these latter differences tends to increase with higher feedback intensity, due to the enhanced stochasticity discussed above.
For the sake of clarity, the results presented in the following refer exclusively to rapidly rotating BHs ($a = 0.99$).

\begin{figure}
    \centering
    \includegraphics[width=\linewidth]{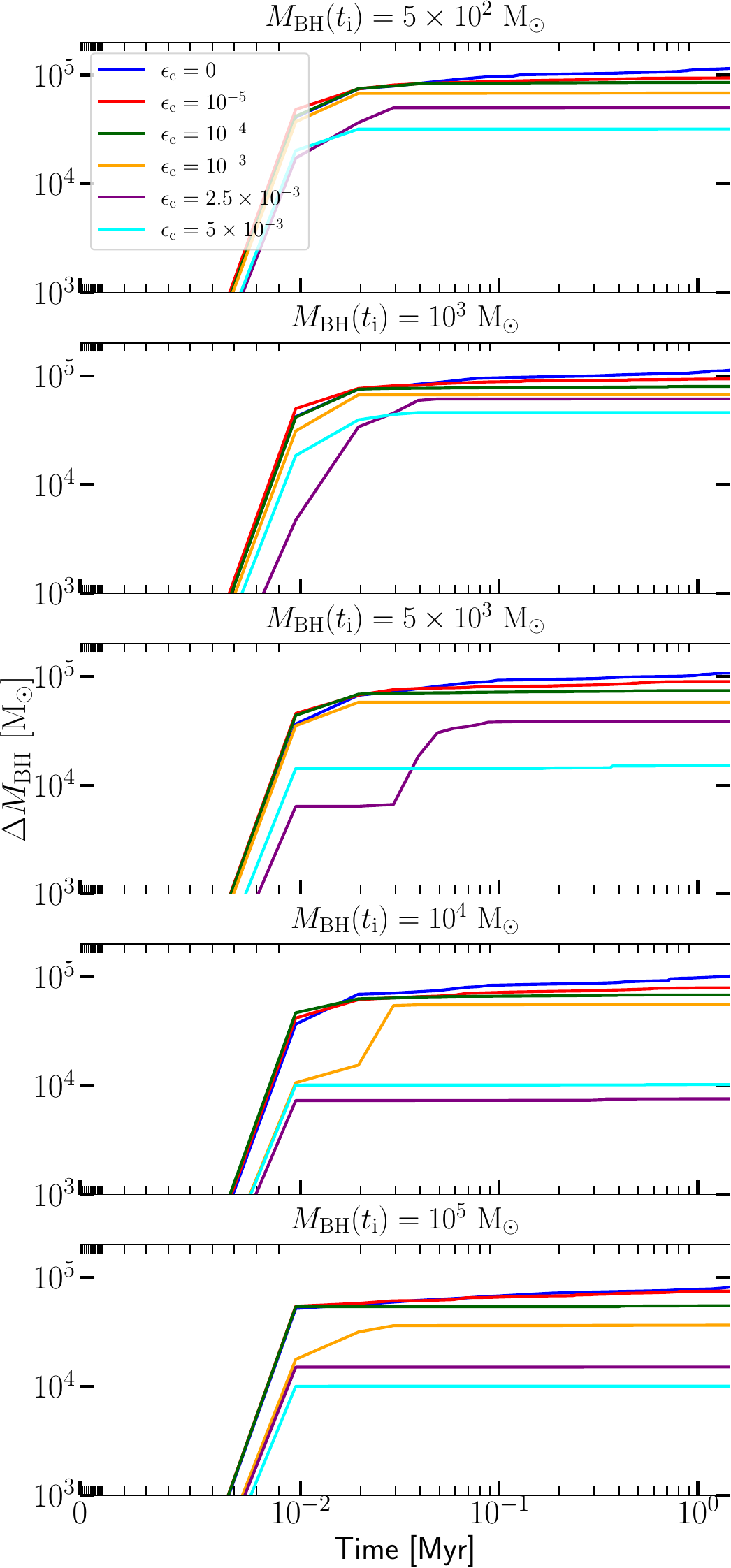}
    \caption{Evolution of the accreted mass $\Delta M_{\rm BH} \equiv M_{\rm BH}(t)-M_{\rm BH}(t_{\rm i})$ for all the tested BH seeds with $a = 0.99$ and different coupling efficiencies. Each plot refers to a different initial mass.
    Different colours indicate the coupling efficiency $\epsilon_{\rm c}$. The same symmetric logarithmic scale and linear threshold of Fig.~\ref{fig:bhmass_evol_eps} are adopted here.}
    \label{fig:bhmass_evol_diff}
\end{figure}
The effect of feedback becomes even more evident in Fig.~\ref{fig:bhmass_evol_diff}, wherein we show the evolution of the accreted mass for our entire suite of runs. The final values of $\Delta M_{\rm BH} \equiv M_{\rm BH}(t)-M_{\rm BH}(t_{\rm i})$ have an inverse dependence on $\epsilon_{\rm c}$ for all the different $M_{\rm BH}(t_{\rm i})$, with the maximum $M_{\rm BH}(t_{\rm f})$ being obviously reached in the $\epsilon_{\rm c} = 0$ cases and decreasing depending on how much energy is redistributed to the surrounding gas. An exception is the case of $M_{\rm BH}(t_{\rm i}) = 10^4~\msun$, in which the growth is slightly larger for $\epsilon_{\rm c} = 5 \times 10^{-3}$ compared to $\epsilon_{\rm c} = 2.5 \times 10^{-3}$, due to the chaotic conditions of accretion, as both cases exhibit high coupling efficiencies.
The BH feedback is relevant, limiting BH growth by up to 90\% (as in the case with $M_{ \rm BH}(t_{\rm i}) = 10^4~\msun$) when the maximum coupling efficiency $\epsilon_{\rm c} = 5 \times 10^{-3}$ is adopted, compared to the case in which feedback is completely absent ($\epsilon_{\rm c} = 0$).

In the following, we restrict our analysis to three representative feedback cases -- $\epsilon_{\rm c} = 0$, $10^{-3}$, and $5 \times 10^{-3}$ -- corresponding to no-, intermediate-, and high-feedback regimes, respectively. This selection is made for clarity in presenting the results, without any loss of generality or information, as evidenced by the trends shown in the previous plots. 

\begin{figure*}
    \centering
    \includegraphics[width=1\linewidth]{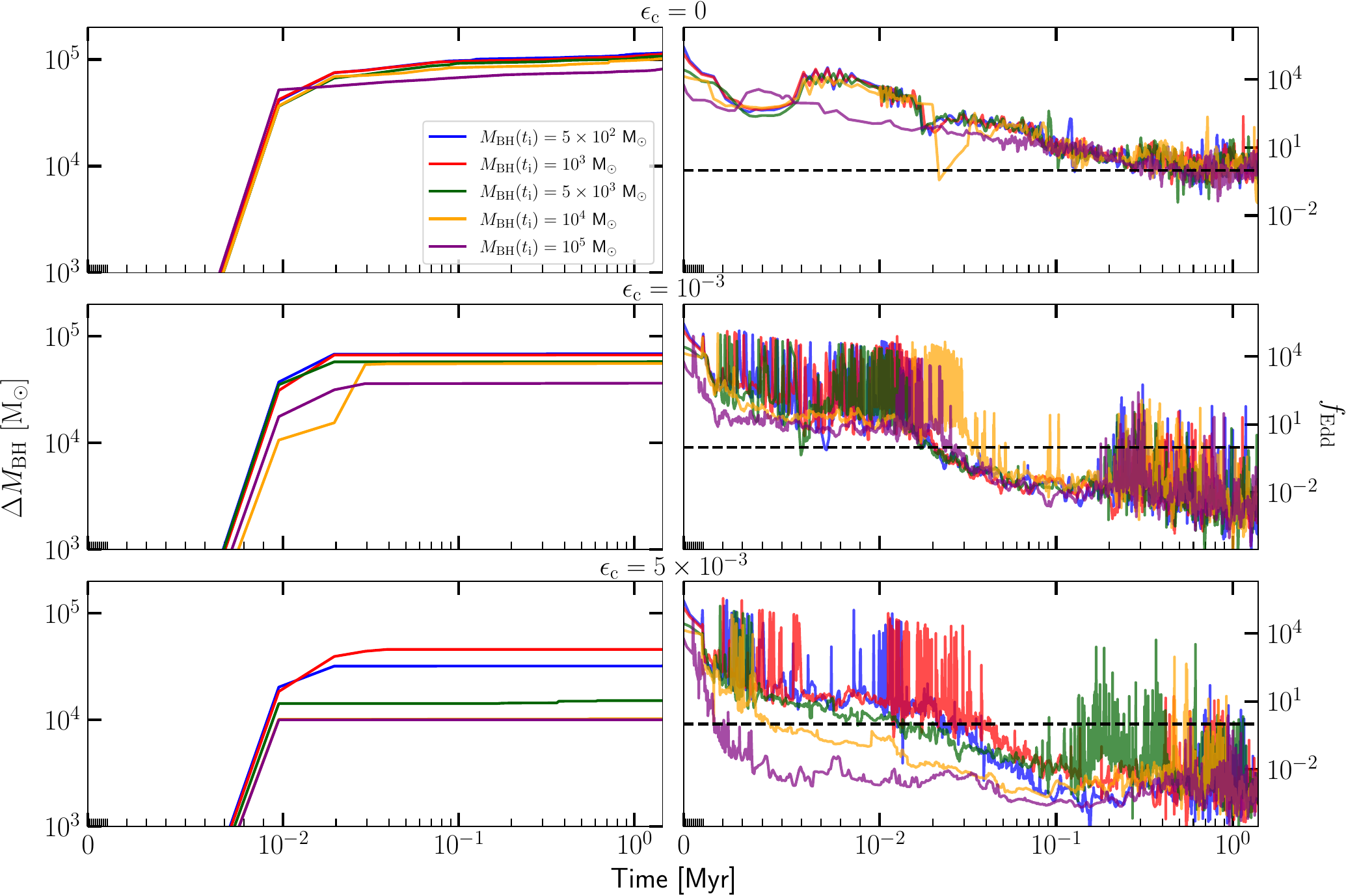}
    \caption{Accreted mass $\Delta M_{\rm BH}~[{\rm M_{\sun}}]$ (left-hand column) and Eddington ratio $f_{\rm Edd} \equiv \dot{M}_{\rm acc} / \dot{M}_{\rm Edd}$ (right-hand column) evolution grouped by feedback intensity for three regimes -- $\epsilon_{\rm c} = 0$, $10^{-3}$, and $5 \times 10^{-3}$ -- from top to bottom, assuming a spin of $a = 0.99$. The Eddington limit is marked by the dashed black line. The same colour-coding and symmetric logarithmic $x$-axis as in Fig.~\ref{fig:bhmass_evol_eps} are used throughout. In the bottommost left-hand panel, the yellow curve is nearly indistinguishable from the purple one, as the two are almost perfectly overlapped.}
    \label{fig:massdiff_fedd}
\end{figure*}

Fig.~\ref{fig:massdiff_fedd} shows the same accreted mass from Fig.~\ref{fig:bhmass_evol_diff}, along with the corresponding Eddington ratio, $f_{\rm Edd} \equiv \dot{M}_{\rm acc} / \dot{M}_{\rm Edd}$, grouped by $\epsilon_{\rm c}$ for the three feedback regimes considered.
Analogously to what is shown in Fig.s~\ref{fig:bhmass_evol_eps} and \ref{fig:bhmass_evol_diff}, the initial rise in $\Delta M_{\rm BH}$ (left-hand column in Fig.~\ref{fig:massdiff_fedd}) is a common feature across all panels. This early growth phase is enabled by a sharp peak in $f_{\rm Edd}$ (right-hand column), briefly exceeding the Eddington limit by over five orders of magnitude at the very beginning of each simulation, as the gas reaches high central densities before feedback becomes effective. These conditions arise from the specific environmental setup adopted in this study, in which the host galaxy is gas-rich, quiescent, and has not undergone any prior SF before the onset of BH accretion.
We note that, although such an environment is clearly idealised -- as required for the type of study conducted -- it nonetheless provides a valuable laboratory for testing the SE accretion regime. The properties of high-redshift galaxies can give rise to similar conditions, even if only briefly, which is sufficient to trigger such intense accretion episodes.
For example, even if previous SF episodes had occurred, our setup is intended to investigate the early phases of bursty SF, where individual SF events are separated by timescales longer than the characteristic duration of a single burst \citep[see e.g.][]{Iyer_et_al_2019, Flores_Velazquez_et_al_2021, Dome_et_al_2024}. In this regime, any pre-existing stellar component would effectively behave as a diffuse background and would not materially influence the system’s dynamical evolution.
After this initial phase, $f_{\rm Edd}$ never reaches the same extreme values again, but it continues to shape the growth of the BHs by progressively building differences in accreted mass of the various runs that depend primarily on the feedback intensity.

Following the initial growth phase, common to all feedback intensities, the accreted mass saturates in the no-feedback case for all simulated seeds. This occurs because nearly all the available gas is rapidly accreted without resistance, as also evidenced by the absence of the ``breathing'' mode in the $f_{\rm Edd}$ plot -- fluctuations that are instead clearly visible in the other feedback regimes.

As the feedback intensity increases, the evolutionary tracks begin to diverge, and successive accretion episodes become generally more significant and are regulated by the energy deposited to the surrounding gas.
The BH `breathing' also becomes stronger and more chaotic for the less massive seeds, reflecting the increasingly bursty nature of the accretion. In the most extreme case with $\epsilon_{\rm c}=5 \times 10^{-3}$, the massive $10^4$--$10^5~\msun$ BHs nearly quench their growth almost immediately, with $f_{\rm Edd}$ dropping to negligible values as the initial feedback-driven outflow succeeds in depleting the nuclear region of gas.
As noted before, for all the feedback regimes, the swallowed gas mass is generally larger for the less massive seeds, and this difference becomes more pronounced as the feedback strength increases. In the high-feedback regime, the cavity inflated by feedback is so effective that the BHs with $M_{\rm BH}(t_{\rm i}) = 10^4$ and $10^5~\msun$ accrete almost exactly the same amount of gas during the first few thousand years, with no further significant accretion episodes throughout the remainder of the simulation.
Interestingly, although to a very limited extent, the tendency for the swallowed gas mass to be larger in less massive seeds is also observed in the no-feedback case.
This may be due to slight differences in the initial equilibrium conditions.\footnote{The discs are relaxed into equilibrium individually for each seed mass.} However, the very initial trend shows that more massive seeds grow slightly faster, as the BHL accretion rate scales with the square of the BH mass ($\dot{M}_{\rm BHL} \propto M_{\rm BH}^2$).

The SE phase ends and $\dot{M}_{\rm acc}$ drops well below the Eddington threshold within the first 100~kyr when feedback is present, whereas it lasts significantly longer -- by about an order of magnitude -- in the no-feedback case.
Apart from the case with $\epsilon_{\rm c} = 5 \times 10^{-3}$ and $M_{\rm BH}(t_{\rm i}) = 10^3~\msun$ -- experiencing a non-negligible accretion event after the initial growth -- a higher coupling efficiency leads, in general, to an earlier transition to sub-Eddington accretion.
This pattern is expected, as sustaining SE accretion becomes increasingly difficult: while $\dot{M}_{\rm Edd}$ grows linearly with $M_{\rm BH}$, the feedback energy -- being typically proportional to $\dot{M}_{\rm BHL}$ and thus to $M_{\rm BH}^2$ -- becomes significantly more effective for more massive BHs.

In summary, it is clear that the Eddington limit is always significantly exceeded at least during the initial 10--100~kyr of evolution.
\begin{figure}
    \centering
    \includegraphics[width=1\linewidth]{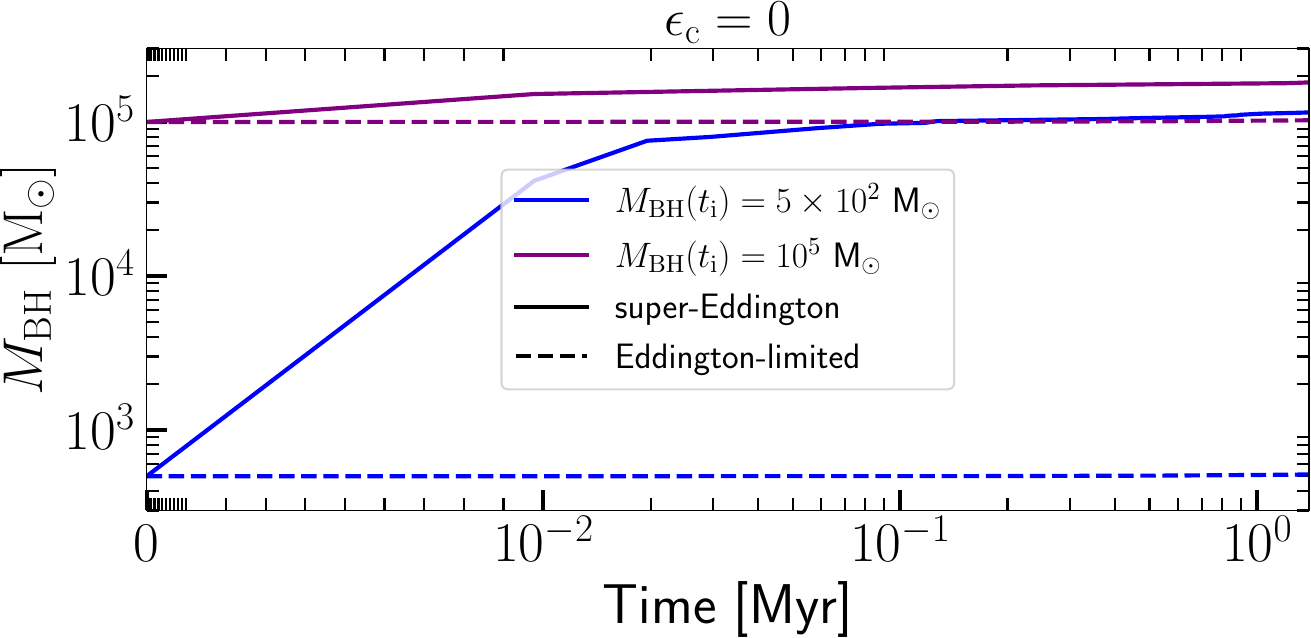}
    \caption{Evolution of the BH mass for rapidly rotating seeds ($a = 0.99$) in the absence of feedback ($\epsilon_{\rm c} = 0$). Solid lines show the SE runs for $M_{\rm BH}(t_{\rm i}) = 5 \times 10^2$ (blue) and $10^5~\mathrm{M}_{\sun}$ (purple), whereas dashed lines indicate the corresponding Eddington-limited cases, shown for comparison with a fixed radiative efficiency of $\epsilon_{\rm r} = 0.1$.}
    \label{fig:comparison_subEdd}
\end{figure}
In Fig.~\ref{fig:comparison_subEdd}, the mass evolution is compared with a scenario in which the Eddington limit (dashed lines) is enforced with a fixed $\epsilon_{\rm r} = 0.1$.
The cases shown correspond to the least ($5 \times 10^{2}{\rm M_{\sun}}$, blue) and most ($10^{5}{\rm M_{\sun}}$, purple) massive BHs in the no-feedback regime, chosen to maximise accretion.
The accretion limited to $\dot{M}_{\rm Edd}$ prevents the BH from growing by more than 3\% for the lightest seed and 2.7\% for the most massive one, in agreement with an exponential growth governed by an $e$-folding time of approximately 45~Myr.
As expected, the accretion of any seed mass over this timescale is negligible under standard accretion mechanisms. The focus therefore shifts entirely to the seed formation process, which determines almost the entire final mass of the BH (as opposed to the SE case), even after about 1.5~Myr of evolution in a gas-rich galaxy without any supernova explosions.
We note that, as we later show, the gas is steadily depleted by the SF processes, which -- together with stellar feedback -- can become a significant limiting factor when accretion proceeds this slowly.

\subsection{Stalled growth}
\label{subsec:stalled_growth}

As shown previously, all BHs undergo rapid early growth during the first few thousand years, after which their growth nearly saturates.
The main reason is that the gas in the vicinity of the BH is almost completely depleted by accretion and by the SF process.

\begin{figure*}
    \centering
\begin{tikzpicture}
  \node[anchor=south west, inner sep=0] (img) at (0,0)
    {\includegraphics[width=\linewidth]{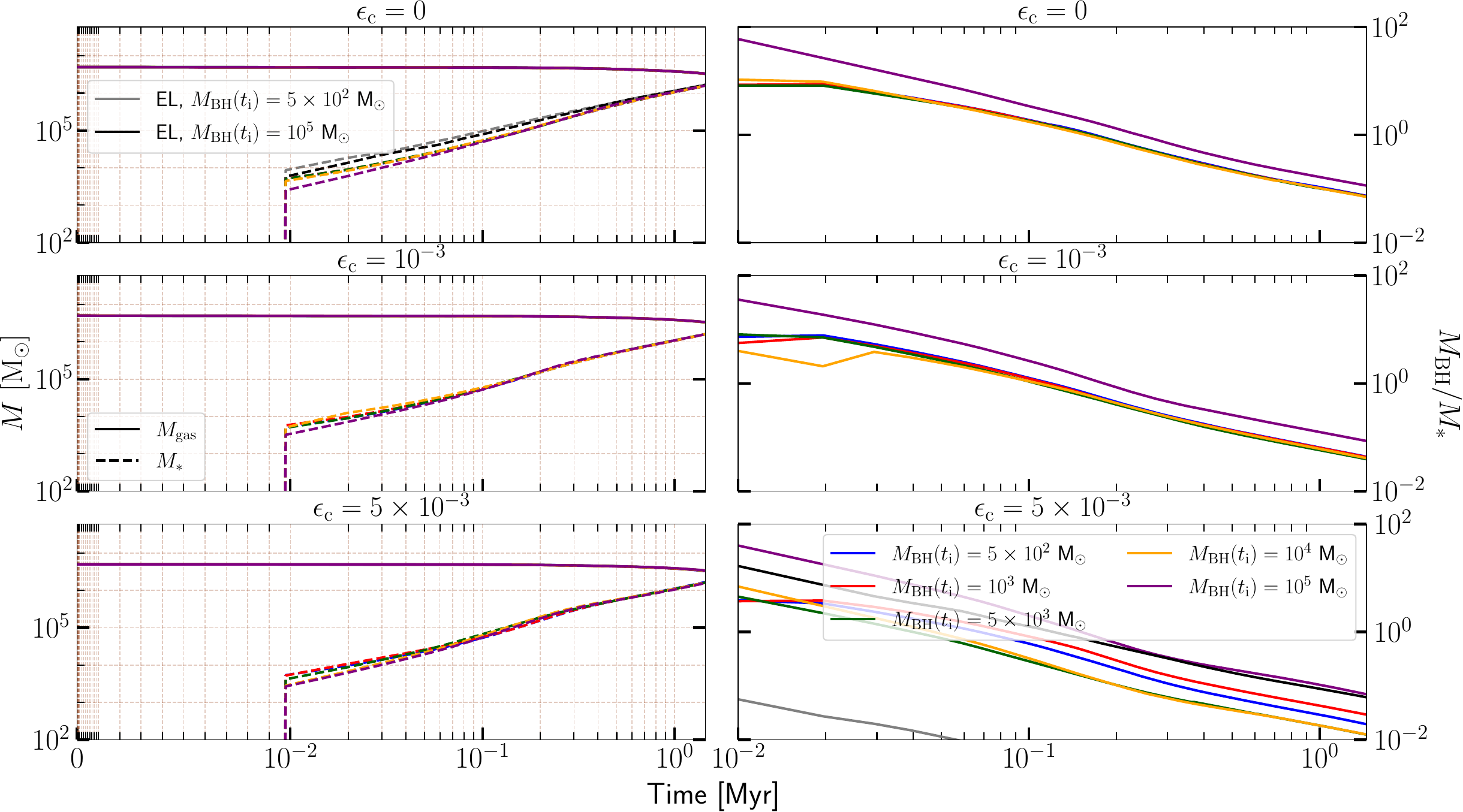}};
  \begin{scope}[x={(img.south east)}, y={(img.north west)}]
    \coordinate (rSW) at (0.476, 0.884); 
    \coordinate (rNE) at (0.482, 0.901); 
    \coordinate (rSE) at (0.488, 0.88); 
    \draw[Sienna, thick] (rSW) rectangle (rNE);
    \node[anchor=north east, draw=Sienna, line width=0.6pt, fill=white, inner sep=0pt] (inset) at (0.93, 0.65)
      {\includegraphics[width=4.5cm]{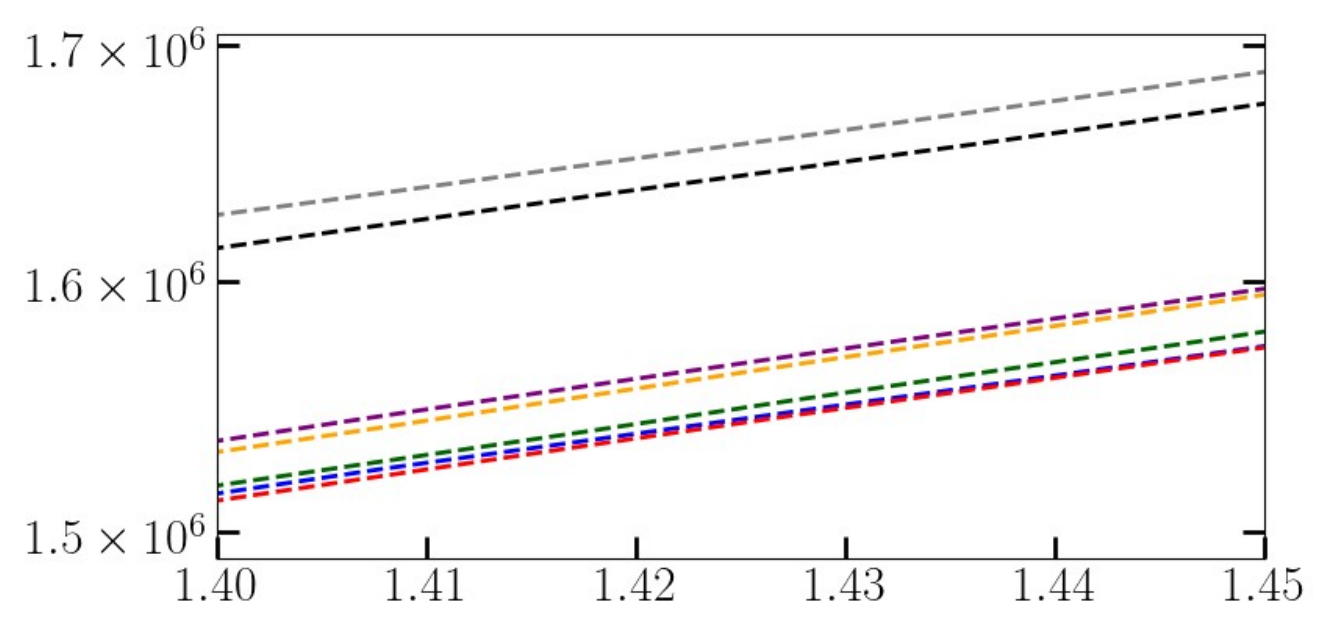}};
    \draw[Sienna, dashed] (rSE) -- (inset.north east);
    \draw[Sienna, dashed] (rSW) -- (inset.north west);
  \end{scope}
\end{tikzpicture}
    \caption{Left-hand panels: Evolution of the total gas mass (solid lines) and stellar mass (dashed lines) within the entire disc for the three feedback cases under analysis (i.e. $\epsilon_{\rm c} = 0$, $10^{-3}$, and $5 \times 10^{-3}$, from top to bottom). The inset magnifies the terminal part of the stellar evolution lines in the top-left panel (the no-feedback case), where the curves overlap. The right-hand panels show the evolution of the $M_{\rm BH}/M_{*}$ ratio. Different colours indicate the various initial BH masses, following the same colour scheme used in Fig.~\ref{fig:bhmass_evol_eps}. For reference, the Eddington-limited (`EL' in the figure) cases with $M_{\rm BH}(t_{\rm i})=5\times10^2$ and $M_{\rm BH}(t_{\rm i})=10^5$ and $\epsilon_{\rm c} = 0$ are included in the corresponding panel. The same symmetric logarithmic scale and linear threshold of Fig.~\ref{fig:bhmass_evol_eps} are applied to the $x$-axis in the left-hand column, whereas a full logarithmic scale is used in the right-hand column, with the origin set at $t = 10^{-2}$~Myr, approximately corresponding to the onset of the first SF episodes.}
    \label{fig:gasnstar_total}
\end{figure*}

In the left-hand panel of Fig.~\ref{fig:gasnstar_total}, we report the evolution of the total gas and stellar mass in the simulated discs for the three feedback instances of interest ($\epsilon_{\rm c} = 0$, $10^{-3}$, and $5\times10^{-3}$).
The SF starts in all cases at approximately $10^{-2}$~Myr, mostly due to the temporal discretisation of the snapshots from which we retrieve information on the stellar mass.\footnote{It should be noted that our SF prescription produces stars on the same timescale as the snapshot interval, i.e. every $\sim$$10^4$~yr.}
From that point onwards, the total stellar mass increases monotonically with time and follows a similar evolution across all runs, showing only small differences depending on $M_{\rm BH}(t_{\rm i})$ and $\epsilon_{\rm c}$.  However, these differences show no clear trend -- for instance, the intermediate-feedback case ($\epsilon_{\rm c} = 10^{-3}$) results in the early highest stellar masses regardless of $M_{\rm BH}(t_{\rm i})$, pointing to a possible sweet spot in BH feedback, where the injected energy is not strong enough to significantly quench SF, yet sufficient to suppress BH growth and retain more gas available to form stars -- possibly even triggering it through shock-induced compression \citep[e.g.][]{Zubovas_et_al_2013, Shin_et_al_2019, Zana_et_al_2022a}. 
Some of these effects are also evident in the total stellar mass $M_{*}(t_{\rm f})$ at the end of the simulations, which tends to increase with stronger feedback, further suggesting a potentially positive role of BH feedback in enhancing SF.
In the sub-Eddington cases, for $\epsilon_{\rm c} = 0$, the stellar evolution is qualitatively similar but shows a consistently higher mass by some $10^{5}~\msun$. This is expected, due to the fact that the amount of gas in the densest regions of the disc remains available for SF, as the BH growth is minimal.
Given that the SE case with $\epsilon_{\rm c}=0$ produces fewer stars than its sub-Eddington counterpart -- initially by nearly a factor of two -- we can confirm that gas availability is the only dominant factor in this context: SE accretion depletes the gas in the central region at a much faster rate than in the sub-Eddington case.

In the right-hand panels of Fig.~\ref{fig:gasnstar_total}, the mass ratio $M_{\rm BH}/M_{*}$ decreases steadily over time as stars are formed.
The different trends in the ratio reflect features similar to those observed in the evolution of the BH mass (see Fig.~\ref{fig:bhmass_evol_eps}), such as the increasing level of stochasticity driven by feedback.
The BHs are overmassive with respect to the stellar component for all initial masses in the SE cases, in accordance with JWST data \citep[e.g.][]{Harikane_et_al_2023, Maiolino_et_al_2024}. Even if there is some dependence on the adopted initial conditions, as no pre-existing stars were present when the BHs started accreting, and BHs become less overmassive as the stellar component builds up, it is noteworthy that such behaviour is strictly linked to the SE case. In contrast, in the sub-Eddington case, the BHs appear overmassive only for the most massive seeds ($M_{\rm BH}=10^{5}\,\msun$ in this analysis).

The total gas mass, on the other hand, decreases from the beginning of the run and settles around $10^{6}\,\msun$ across all $M_{\rm BH}(t_{\rm i})$.\footnote{We note that the logarithmic scale hampers a clear visual comparison of the small residual differences at those values.}
By the end of the simulation, the gas mass is only halved, and the global stellar mass keeps increasing, with no sign of saturation. This indicates that the BH has exhausted the gas only in its immediate surroundings.

In conclusion, it is necessary to examine a smaller region around the BH to shed light on the mechanisms at play.

\begin{figure}
    \centering
    \includegraphics[width=1\linewidth]{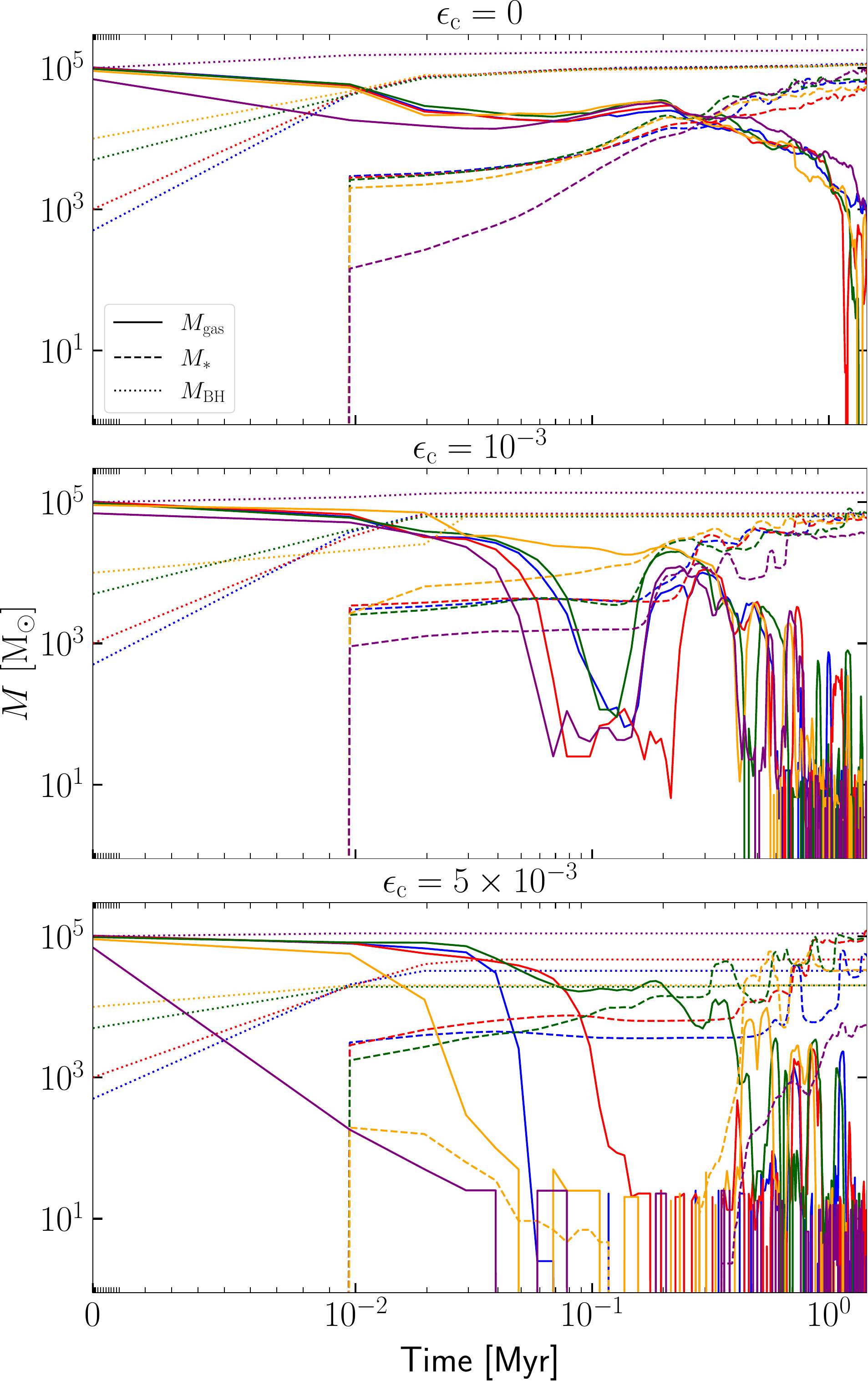}
    \caption{Same as in the left-hand panels of Fig.~\ref{fig:gasnstar_total} but, for the same three values of $\epsilon_{\rm c}$ and colour-coding, the evolution of the gas mass (solid lines), stellar mass (dashed lines), and $M_{\rm BH}$ (dotted lines) are computed within 1~pc from the BH.}
    \label{fig:gasnstar_1}
\end{figure}

Such close-up is provided by Fig.~\ref{fig:gasnstar_1}, where we show the evolution of the gas, star, and BH mass for the same feedback regimes of Fig.~\ref{fig:gasnstar_total}, but within a sphere of radius 1~pc, which initially encloses a gas mass comparable to the total mass accreted by the most rapidly growing BHs ($\sim$$10^5~\msun$).
This smaller region is selected to be close enough to the BH to probe its feeding process, but not too close to be overly sensitive to fluctuations, which arise both from the inflow of material into the sphere and from the BH’s own migration. The latter is consistently more pronounced in the high-feedback case, as shown in the bottom panel.

In all cases, while the stellar and BH masses continue to grow -- albeit at decreasing rates -- the gas mass experiences a noticeable decline around $\sim$$0.1~\mathrm{Myr}$, with its onset occurring earlier for stronger feedback cases, as expected due to the increased strength of outflows.
We observed no coherent dependence on the initial BH mass; however, in nearly all cases, the 1-pc sphere becomes entirely depleted of gas. From that point onwards, only the migrating BH is able to encounter additional fuel while orbiting through the disc, enabling occasional minor accretion episodes.

In general, we find that SF does consume a portion of the gas in the vicinity of the BH. However, the timescales associated with accretion are significantly shorter, and as a result, the initial accretion episodes are only marginally affected.

\begin{figure*}
    \centering
    \includegraphics[width=1\linewidth]{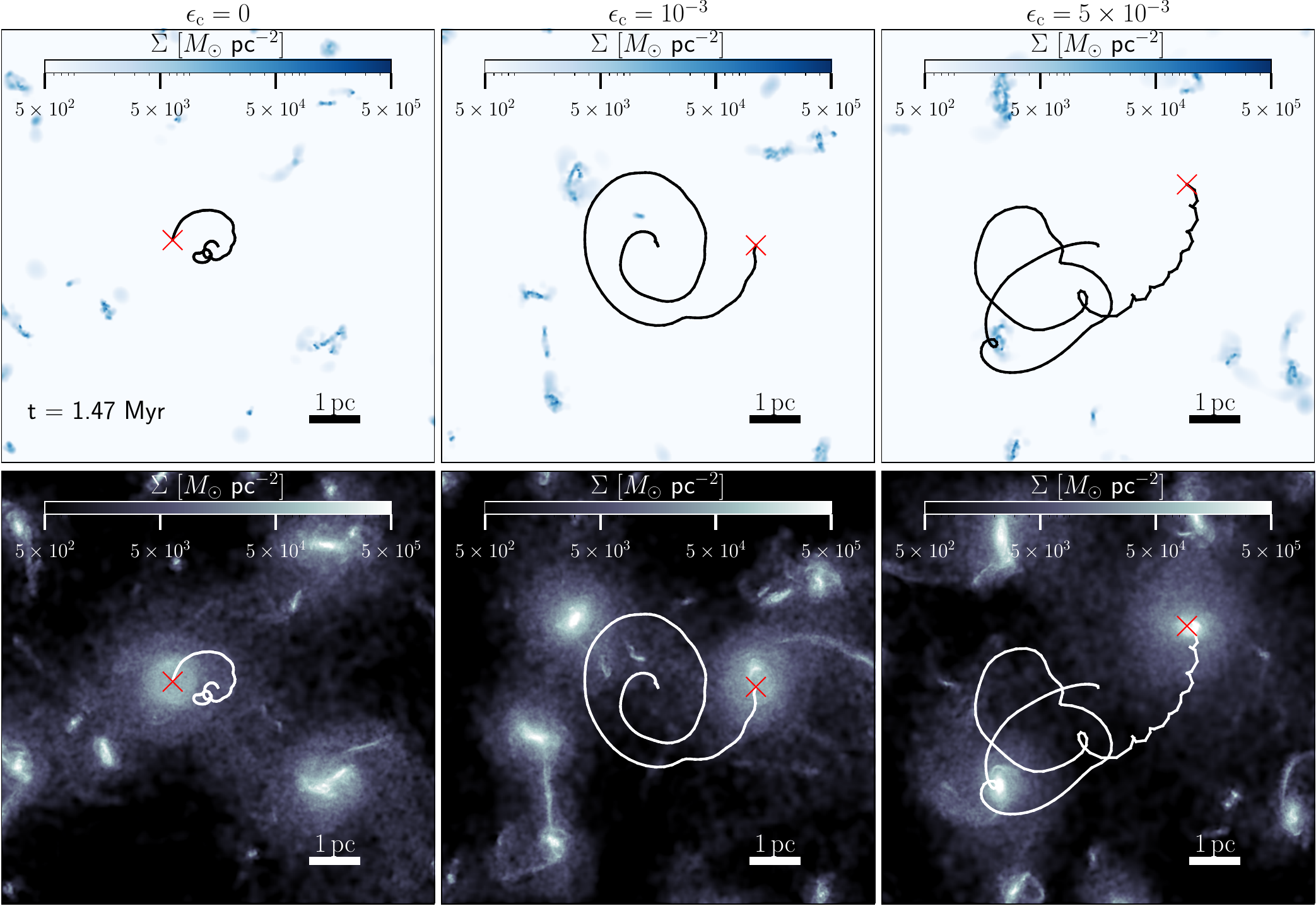}
    \caption{Surface density maps ($\Sigma$) of the gas (top row) and stellar (bottom row) components at the final snapshots of the simulations with $M_{\rm BH}(t_{\rm i})=5 \times 10^3~\msun$, shown for three different feedback regimes of interest. The BH orbit is traced in black (top row) and white (bottom row), with the current BH position marked by a red cross.}
    \label{fig:6maps_150}
\end{figure*}
In Fig.~\ref{fig:6maps_150}, we present as a representative example the final snapshot of the simulation with $M_{\rm BH}(t_{\rm i})=5 \times 10^3~\mathrm{M}_{\sun}$, displaying both the gas (top row) and stellar (bottom row) surface density maps in the three feedback regimes we focus on. The whole migration path of the BH is clearly marked in each panel.
Each BH follows an orbit that brings it to a distance of at least $\sim$1~pc from its original position. However, the stronger the feedback is, the more scattered and irregular the BH trajectory appears (see Fig.~\ref{fig:6maps_10} for an earlier stage of the same simulations).
As can be evinced from the stellar components, the BH is consistently embedded within a massive stellar cluster. This cluster forms and grows progressively from the gas orbiting the BH itself, as well as through continuous mergers with other clusters in the disc. Its gravitational influence significantly affects the BH's orbital evolution, while its own spatial distribution is strongly modulated by the intensity of the feedback.
When $\epsilon_{\rm c} = 0$, the absence of feedback allows SF to proceed undisturbed in the central region. This leads to the formation of a dense central cluster that helps anchor the BH in place through its gravitational pull. Conversely, in the presence of stronger feedback, SF happens at larger distances. As a result, the formation of various offset clusters causes a significant scatter in the BH path, as previously seen in other works \citep[][]{Tamburello_et_al_2017, SouzaLima_et_al_2017}.

On the other hand, the top row clearly shows that the remaining gas is very sparse, with only a few gas clouds exceeding surface densities of about $10^4~\msun$~pc$^{-2}$, and certainly not in the vicinity of the BH, where the gas has already been accreted at this stage, as previously seen in Fig.~\ref{fig:gasnstar_1}.
The BH's orbiting remains the only mechanism capable of triggering new, albeit limited, episodes of accretion.
In conclusion, the effect of SF is particularly complex, as it not only provides a competing mechanism that consumes gas in the vicinity of the BH, but also influences its orbital motion.

To better isolate this effect, we re-ran some of the simulations presented here without including SF. Specifically, we performed a set of simulations with the same parameter set as that of our fiducial suite, with the sole exception of disabling SF (hereafter referred to as noSF). In addition, we carried out an analogous set of simulations in which we also raised the temperature floor from $44$~K to $3 \times 10^3$~K (hereafter named Tf). This was introduced to provide an additional pressure support to the gas and to suppress gravitational clumping, which was still present even in the absence of SF. The goal was to keep more gas available for BH accretion by mitigating the collapse of gas into dense structures far from the BH itself.
\begin{figure}
    \centering
    \includegraphics[width=\linewidth]{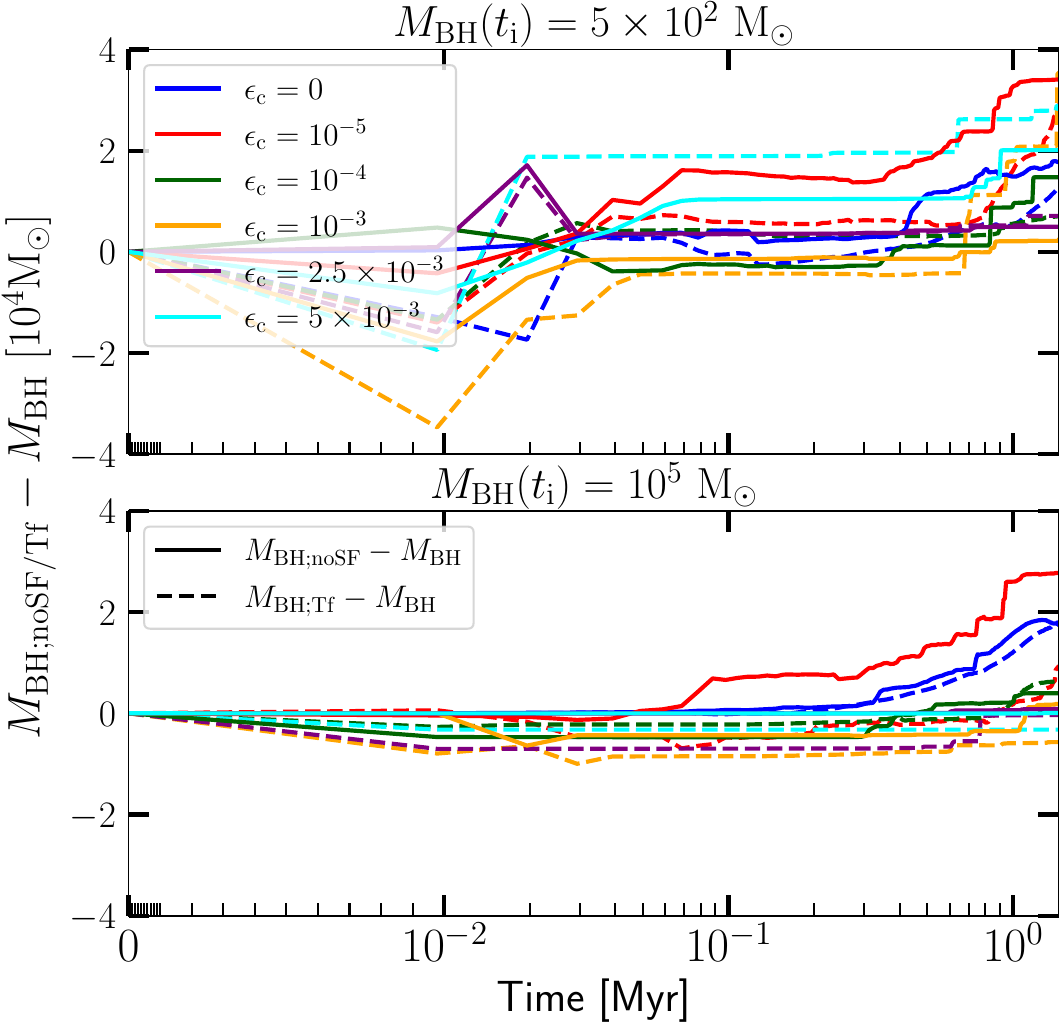}
    \caption{Mass difference between the BH masses $M_{\rm BH; noSF}$ -- obtained from the simulations without SF (noSF, solid lines) -- or $M_{\rm BH; Tf}$ -- from simulations without SF and with an increased temperature floor (Tf, dashed lines) -- and the reference masses $M_{\rm BH}$. The same $x$-axis and colour-coding as in Fig.~\ref{fig:bhmass_evol_diff} are used. The upper panel refers to the suite with $M_{\rm BH}(t_{\rm i})=5\times10^2~{\rm M_{\sun}}$, whereas the lower panel refers to the suite with $M_{\rm BH}(t_{\rm i})=10^5~{\rm M_{\sun}}$.}
    \label{fig:massgain_vs_noSF}
\end{figure}

In Fig.~\ref{fig:massgain_vs_noSF}, we show the evolution of the difference between the BH mass in the simulations noSF (solid lines) and Tf (dashed lines) and their counterpart in our fiducial suite $M_{\rm BH}$, for all feedback regimes and for two representative initial masses at the opposite ends of the explored mass range, namely $5\times10^2~{\rm M_{\sun}}$ and $10^5~{\rm M_{\sun}}$.

Apart from an initial phase, during which the absence of SF results in a complex and chaotic accretion behaviour, the general trend across all simulations shows a higher final $M_{\rm BH}$ in both suites without SF, with two minor exceptions occurring in the cases with $\epsilon_{\rm c}=10^{-3}$ and $5\times10^{-3}$ in Tf for $M_{\rm BH}(t_{\rm i})=10^5~\msun$.
The final mass difference in the noSF suite reaches $3.4\times10^{4}~\msun $ for the case with $\epsilon_{\rm c}=10^{-5}$ and $M_{\rm BH}=5\times10^2~{\rm M_{\sun}}$, and $2.7\times10^{4}~\msun$ for $M_{\rm BH}=10^5~{\rm M_{\sun}}$.
In the Tf suite, the maximum differences are $3.5\times10^{4}~\msun$ for $\epsilon_{\rm c}=10^{-3}$ and $M_{\rm BH}=5\times10^2~{\rm M_{\sun}}$, and only $1.8\times10^{4}~\msun$ for $\epsilon_{\rm c}=0$ and $M_{\rm BH}=10^5~{\rm M_{\sun}}$. 

In the noSF suite, the differences at the end of the runs are consistently large across different initial BH masses in the cases with no or low feedback.
Even in the absence of SF, gas clumps still form and influence the orbital evolution of the BH. Additionally, anisotropies in the outflowing gas at the beginning of the simulation induce a kick which contributes to the initial displacement of the BH. As a result, even here feedback increases the amplitude of BH orbits.
When feedback is weak, independently of the initial mass of the BH, its motion remains limited: it does not wander significantly, and all the formed gas clumps tend to coalesce in the central region, binding to the BH. The compact object then orbits together with this central structure, which continuously supplies it with gas. Since SF is absent, this inflow is uncontested, allowing for larger accretion.
Conversely, as the feedback intensity increases, interactions with dense gas clumps become more stochastic, and accretion becomes less efficient.

In the Tf suite, the behaviour of accretion is different, as the formation of dense clumps is strongly suppressed (though not completely avoided), and gas remains pervasive and much more evenly distributed throughout the disc.
In this case, the displacement of the BH is primarily driven by anisotropies in the surrounding gas density field, causing a mildly stochastic accretion. When feedback is strong, the absence of massive gas (or stellar) clumps embedding the BH poses no constraint on its motion, allowing it to wander freely. In contrast, in low-feedback regimes, the BH tends to remain near the centre, as previously observed (see Fig.~\ref{fig:dist_vs_noSF}).
The lack of dense gas clumps that sustain BH accretion makes feedback far more decisive in regulating inflows. As a result, accretion becomes highly stochastic, driven by the BH’s orbit in the strong feedback regime, and by the formation of gas structures -- such as spiral arms funnelling material towards the BH -- in the low-feedback cases.

Overall, and with some exceptions, BH growth tends to be larger in the absence of SF. However, even in these cases it typically saturates eventually, unless the BH continues to be scattered across the disc, reaching very large distances from the centre. This occurs only in less physical scenarios (i.e. without SF and with an increased temperature floor) and, even in those cases, the final mass is often not significantly higher than that of the fiducial case within the simulated time frame (e.g. for $M_{\rm BH}=5\times10^2~{\rm M_{\sun}}$ with $\epsilon_{\rm c}=5\times10^{-3}$ and $M_{\rm BH}=10^5~{\rm M_{\sun}}$ with $\epsilon_{\rm c} \geq 10^{-3}$).
The case with $M_{\rm BH}=5\times10^2~{\rm M_{\sun}}$ and $\epsilon_{\rm c}=10^{-5}$ in Tf is particularly interesting, as it appears to be the only one that did not reach saturation by the end of the simulation. At that point, it is already 31\% more massive than its fiducial counterpart. This behaviour is driven by a random gas overdensity that is funnelled towards the disc centre, where the BH still resides. This further suggests that accretion could easily resume as soon as new material becomes available.

\section{Summary and discussion}
\label{sec:discussion}

We have observed that SE accretion acts as a game-changer, as it significantly reduces the relevance of both the BH seed formation channel and the accretion duty cycle. This is because large amounts of gas (up to $10^5~\msun$) can be efficiently accreted by the BH seed at any time over a short timescale of order $10^3$--$10^4$~yr, provided that the appropriate background conditions are in place.
As a matter of fact, our lightest seeds -- $M_{\rm BH}(t_{\rm i})=5\times10^2~{\rm M_{\sun}}$ -- grew the most.
Such rapid accretion phases with short duty cycles are consistent with a wide range of independent studies, including hydrodynamical simulations \citep[e.g.][]{Lupi_et_al_2024a} and those discussed below, semi-analytical models such as CAT \citep{Trinca_et_al_2022}, and observational constraints from $z \sim 6$ quasar proximity zones \citep[e.g.][]{Satyavolu_et_al_2023}.
Similar trends have also been found in the recent works by \citet{Mehta_et_al_2024} and \citet{Gordon_et_al_2025}, who report comparably rapid though somewhat less massive episodes of SE growth lasting well below a megayear. A direct quantitative comparison, however, is not straightforward, as these two studies adopted very different numerical setups and environmental conditions.

In particular, \citet{Gordon_et_al_2025} employed a rather simplified SE accretion prescription in which the classical BHL model was used without enforcing the Eddington limit and the radiative efficiency was kept fixed throughout each accretion episode (they tested only two values: 0.01 and 0.10). Moreover, they applied their SE accretion prescription in a cosmological zoom-in simulation in which most of the relevant sub-grid physics was switched off until the accretion phase began and with SF never activated at any point. Despite having different accretion recipes (but similar star-forming environments, with no SF at the onset of BH accretion), they obtained results that are broadly consistent with ours: SE accretion phases occur when large amounts of cold gas are available; they are short-lived but highly efficient in growing intermediate-mass seeds ($10^{3}$--$10^{5}\,M_{\odot}$, whose initial values are not tightly controlled due to the cosmological setup); and the feedback efficiency is the key regulator determining how freely SE accretion can proceed.

On the other hand, \citet{Mehta_et_al_2024} performed high-resolution isolated simulations more similar in spirit to our approach, but unlike in our case, they simulated an entire population of BHs within a single run, seeded according to a PopIII-based prescription. The resulting seed masses are all below $10^{3}~\msun$ and exhibit a distribution peaked around $2\times10^{2}~\msun$, which is smaller than (though still broadly consistent with) the lower end of our seed-mass range of $500~\msun$.
Most importantly, their simulations do not include any BH feedback, which severely limits the comparison with our results.
Feedback is a fundamental aspect of our framework, and as we discuss below, it is possible that even our model may underestimate its impact.
The only meaningful conclusion that we can draw from a comparison between these works is that in \citet{Mehta_et_al_2024} they also find that BHs can grow efficiently (up to $\approx 10^{5}\,M_{\odot}$, as in our case) through bursts of SE accretion lasting $\sim$$10^{4}$~years, in the absence of supernova feedback.
In the following, we perform a more detailed comparison with the more closely related work by \citetalias{Sassano_et_al_2023}.

Unlike the initial seed mass, the efficiency of feedback plays a primary role, as it can effectively -- although only partially -- limit the growth of the BH. Its influence is highly non-linear, given that its interplay with the SF process is itself crucial for BH growth. In particular, feedback affects BH evolution by ($i$) expelling gas from the central region, thereby allowing it to be consumed by SF -- which, in a closed system, becomes a key channel for gas removal -- and ($ii$) influencing the BH's migration, both by triggering anisotropies in the local density field and by promoting (or suppressing) the formation of stellar clusters at varying distances from the centre, which act as important gravitational attractors.
In general, feedback tends to enhance the overall dynamical chaos of the system, especially by scattering the BH in the disc.

The BH spin does not have a significant impact on the accretion process. However, we confirm that the total accreted mass is slightly higher in the case of a non-spinning BH compared to a rapidly rotating one.
In conclusion, it is difficult to predict the net impact of feedback beyond zeroth order, even within this highly idealised framework.

More than feedback itself, it is the local conditions of gas density, temperature, and velocity field that determine the magnitude of the accretion rate. In this context, thanks to the SE prescription, no artificial upper cap is imposed on the accretion rate, allowing the vast majority of the mass to be accreted at the beginning of the simulation, before the onset of the first supernovae, which would likely deplete the remaining gas reservoirs left over from SF and BH accretion.

Eventually, accretion is halted or becomes negligible solely because the gas in the very central region of the galactic disc -- the domain effectively accessible to the BH -- is exhausted. The extent of this region can vary significantly depending on how far the BH wanders from the galactic centre and this, in turn, depends on feedback.

These results indicate that, on the one hand, the initial conditions of our suite are crucial in driving such a rapid BH growth; in the real Universe, extremely high gas densities (e.g. $>10^5\,{\rm M}_{\sun}\,\mathrm{pc}^{-3}$ within the central parsecs) are required to support such episodes of intense accretion. On the other hand, they suggest that if similar physical conditions are re-established at later stages of galaxy evolution, for instance via galactic mergers \citep[e.g.][]{Capelo_Dotti_2017} or via secular evolutionary processes \citep[e.g.][]{Fanali_et_al_2015}, a comparable burst of accretion could occur again (see \citealt{Lupi_et_al_2024a} for quantitative examples).

Moreover, we have shown that not only SF acts as a competitive mechanism, since gas that remains available can contribute to BH accretion, but also that the dynamical motion of the BH is profoundly affected by the formation of self-gravitating stellar clusters, which ultimately play an equally important role in shaping the BH accretion history.

Our findings open up a range of scenarios in which the constraints on the epoch of seed formation can be relaxed, since short SE bursts with low duty cycles may still enable BHs to reach the observed masses by the epoch of reionization, as observed by \citet{Trinca_et_al_2023, Trinca_et_al_2024}.

From our analysis, it appears that SE accretion naturally leads to the formation of BHs that are overmassive with respect to the host galaxy stellar mass. This provides a promising pathway to account for the broad population of overmassive BHs recently observed with JWST at high redshift.

It is worth drawing a specific comparison with the work by \citetalias{Sassano_et_al_2023}, as we inherited from them both the earlier version of the initial conditions and the first implementation of the code. In our study, we focused exclusively on the BH feedback prescription proposed by \citet{Madau_et_al_2014}, since it was identified in analysis by \citetalias{Sassano_et_al_2023} as the most efficient in terms of total accreted mass.

Due to suboptimal code performance and the setup of the initial conditions adopted in their work (see Sec.~\ref{sec:simulations}), the authors significantly underestimated the possible accreted mass of a $10^3~\msun$ BH seed in the case with $\epsilon_{\rm c}=10^{-5}$, which they quantified as approximately $2 \times 10^4~\msun$, whereas in their case with $\epsilon_{\rm c}=10^{-3}$, the BH barely grew, increasing its mass by only 70\% after 1~Myr.
By applying the same simple extrapolation model proposed in \citetalias{Sassano_et_al_2023} -- that is, by estimating the number of possible accretion events triggered by gas inflows following galaxy mergers, computed as the ratio between the cosmological time interval between $z = 15$ and $z = 6$ and the free-fall time of the galaxy\footnote{To evaluate the free-fall time, we estimate $\sim$$350~\msun~\mathrm{pc}^{-3}$ to be the total matter density within a cylindrical region of height 2.75~pc and radius 55~pc, corresponding to the parameters of the Mestel disc initialised in our simulation suite. The measurement comes from the final snapshot of the run with average feedback intensity, which we consider the most representative of the entire set.} -- and multiplying such number ($\sim$1000) by the accreted mass measured in our simulations, we obtain a final BH mass of $\sim$$9.5 \times 10^7~\msun$ at $z=6$ in the same runs performed in \citetalias{Sassano_et_al_2023}, i.e. with $M_{\rm BH}(t_{\rm i})= 10^3~\msun$, and $\epsilon_{\rm c} = 10^{-5}$. This value increases to about $1.2 \times 10^8~\msun$ in the no-feedback scenario with an initial seed mass of $M_{\rm BH}(t_{\rm i})=5 \times 10^2~\msun$, which corresponds to the BH that grows the most in our suite. 
In this context, adopting a more massive initial seed proves counterproductive, as it does not lead to higher final BH masses, thereby questioning the advantage of massive seed formation channels in this regime.
The final BH masses we obtain are very close to those of the most massive BHs observed at the epoch of reionization, although they still fall slightly short of the highest observed values. In this context, however, we could also speculate that continuous nearly-Eddington accretion may occur once the host galaxy becomes sufficiently massive, allowing the BHs to reach the upper end of the observed distribution \citep[see e.g.][]{Di_Matteo_et_al_2017, Angel-Alcazar_et_al_2017, Lupi_et_al_2019}.
Finally, if we take the previous estimate of the number of accretion events and multiply it by the duration of each episode, which in our study lies in the range $10$--$100$~kyr, we obtain a duty cycle of $\sim$1.5--15\% over the cosmic period from $z=15$ to $z=6$. These relatively short duty cycles are consistent with recent estimates for high-redshift quasars at $z \approx 6$ based on JWST observations \citep[][]{Pizzati_et_al_2024, Eilers_et_al_2024}.

In any case, this extrapolation model is extremely simplistic, as it is inherently difficult to estimate the actual amount of gas that can reach the central regions of a galaxy. However, through our analysis, we have demonstrated that there is no bottleneck at the level of BH accretion itself, within the scale of our accretion prescription, which coincides with the BH smoothing length (ranging from $10^{-3}$~pc up to a maximum of $\sim 2$~pc).
The problem therefore shifts to understanding the mechanisms driving gas inflows towards the galactic centre, i.e. the so-called angular momentum problem. In this context, it becomes crucial to identify how the gas in the disc, especially at later evolutionary stages, can effectively lose angular momentum and be funnelled into the BH’s sphere of influence.
Possible mechanisms include galaxy mergers, turbulence (either merger-induced or driven by supernova explosions), or the formation of non-axisymmetric stellar structures such as bars or spiral arms \citep[see][for a recent review]{Capelo_et_al_2023}, which are being observed in increasing numbers and at progressively higher redshifts \citep[][]{Zana_et_al_2022b, Costantin_et_al_2023}.

\subsection*{Caveats}

The interpretation of our results requires a proper discussion of the key limitations inherent to our approach.

\paragraph{($i$) Isolated system.}
Our conclusions are drawn within a highly idealised framework in which the galaxy is isolated, experiencing neither cosmological inflows nor tidal interactions with other structures, in tension with the expectation that massive high-redshift quasars live in very dense environments \citep[e.g.][]{Zana_et_al_2023}. On the one hand, this increases the likelihood of rapid, massive accretion, as the disc is initially in equilibrium with the BH and hosts a large reservoir of central gas. On the other hand, the absence of external gas replenishment prevents sustained accretion beyond the first megayear.
In order to achieve the resolution adopted in this work, it is necessary to rely on isolated setups due to obvious limitations in computational resources. However, we note that we are currently working on extending this study to a cosmological context (Caleno et al., in prep.), by employing zoom-in techniques and particle splitting methods. This will allow us to investigate, for the first time at this resolution, the role of galaxy mergers in feeding the central BHs on small scales.
\paragraph{($ii$) Initial conditions.}
Initial conditions -- in particular, the choice of the gas density profile and its velocity field -- represent another key influencing factor. We observed that the initial gas density surrounding the BH seed plays a significant role in determining the magnitude of the earliest accretion phases.
We note, however, that the galaxy is in equilibrium with the BH itself, so there are no unnatural gas inflows triggered solely by the BH presence as a perturbation to the system.
We are currently investigating to what extent this initial configuration affects the final BH mass (Zana et al., in prep.).
We note that also the wandering dynamics of the BHs is in principle dependent on the adopted initial conditions, which set the characteristic extent of their displacement and the maximum distances they can reach.
\paragraph{($iii$) No SF before black hole seeding.}
Analogously, even if our assumption of no SF occurring prior to BH seeding is necessary to reduce the complexity of the problem, it also introduces a further deviation from realistic conditions. 
In real systems, stars would possibly consume part of the gas mass available for BH accretion in the immediate vicinity of the seed, although we do not expect this effect to play a dominant role in the present context. More critically, early stellar feedback could affect the gas density, influencing both subsequent episodes of SF and the spatial distribution of gas relevant for accretion.
\paragraph{($iv$) Finite spatial and mass resolution.}
Focusing more closely on the BH itself, we note that the adopted spatial and mass resolution, while already sufficient to capture the dynamics of the gas in the vicinity of the massive object, are not yet optimal. A possible future improvement could involve further increasing the mass and space resolution below the \citeauthor{Bondi_1952} radius. This would allow us to move beyond the current accretion prescription, which is based on Eq.~\eqref{eq:BHL}, and potentially resolve the actual mass fluxes that feed the BH accretion disc.
\paragraph{($v$) No treatment of gas and black hole angular momentum.}
Our accretion prescription is further limited by the fact that it does not account for the angular momentum of the accreted gas. This simplification may lead to a substantial overestimation of the accretion rate, as all gas gravitationally captured by the BH in our setup is accreted, with the only exception of the fraction of mass-energy released through feedback.
\paragraph{($vi$) No mechanical feedback.}
We note that no feedback is included in \citet{Kao_et_al_2025}, as their work presents the first implementation of the angular momentum evolution model. In contrast, our simulations include radiative thermal feedback, redistributed isotropically according to the coupling constant $\epsilon_{\rm c}$, making our setup more realistic in this regard. However, we acknowledge that our approach still lacks a more sophisticated treatment of feedback, such as the inclusion of mechanical feedback in the form of bipolar outflows or jets from the BH-accretion disc system \citep[][]{Sala_et_al_2021, Bollati_et_al_2023}. This component could significantly affect the density of the surrounding gas and therefore the accretion rate, as demonstrated by \citet{Regan_et_al_2019, Quadri_et_al_2025}.
It is, however, difficult to predict the outcome with certainty, as the suppression of accretion may be confined only to the region where the outflows are launched \citep{Hartwig_et_al_2018}. We therefore expect a range of possible outcomes, depending on the specific spatial and physical conditions involved.

\section{Conclusions}
\label{sec:conclusion}

Through a large suite of high-resolution hydrodynamical simulations, we followed the growth of a BH seed accreting in the centre of a gas-rich protogalaxy at $z \sim 15$. By implementing an SE accretion prescription based on the slim-disc model, we systematically assessed the impact of the initial seed mass, feedback efficiency, BH spin, and SF on the resulting BH growth.
Our results are summarised as follows: 
\begin{enumerate}[label=(\it \roman*)]
    \item The SE accretion is highly efficient. The SE accretion enables the BH to accrete nearly all the available gas in its surroundings, regardless of its initial mass and spin. In our simulations, all BHs accrete a total mass ranging from $10^4$ to $10^5~\msun$ within the first $10^3$--$10^4$~yr, after which the growth nearly saturates due to the absence of any global mechanism to replenish gas in the central region.
    
    \item When a sufficient amount of gas is available, feedback becomes the main factor influencing accretion. It reduces the accreted mass both directly, through the redistribution of energy, and indirectly, by altering the dynamics of the BH through the formation of different stellar distributions. Nonetheless, even in the presence of strong feedback, BHs grow significantly more than in the Eddington-limited case, wherein the final mass of the BH still shows a strong dependence on the seed mass.

    \item When feedback is weak, accretion becomes so efficient and tightly regulated by the initial gas reservoir that the system rapidly loses memory of the seed mass. This behaviour is particularly interesting in the context of BH seeding mechanisms, as it implies that different seed formation channels may converge to similar early growth histories. When feedback is strong, by contrast, the final BH masses do not converge to the same value, yet they still do not show any clear dependence on the initial seed mass, further reinforcing the idea that the early growth phase is largely insensitive to the details of the seeding mechanism.
    
    \item After less than a megayear, none of the BHs experience significant further growth, as the gas in the immediate vicinity has been consumed. While SF competes for gas and slightly reduces BH growth, its effect is subdominant because the majority of the accretion occurs on very short timescales ($\lesssim10^4$~yr).
    
    \item The SE accretion naturally drives the BH to be overmassive relative to the stellar component of its host galaxy due to the rapid timescales over which the mass is accreted.
    
    \item In our setup, SE accretion allows BHs -- regardless of initial mass -- to reach the masses of the most massive BHs observed at high redshift through short efficient bursts (i.e. low duty cycles).
    
\end{enumerate}

\begin{acknowledgements}

The authors are grateful to the anonymous referee for the constructive comments and suggestions, which helped improve the quality of the manuscript.
This work was supported by the EuroHPC Joint Undertaking through the allocation of computing resources under the project EHPC-DEV-2024D04-058, on LEONARDO supercluster.
TZ acknowledges the CINECA award under the ISCRA initiative for providing high-performance computing resources and support on GALILEO supercluster.
TZ, RS, and LG acknowledge support from the EU-Recovery Fund PNRR - National Centre for HPC, Big Data and Quantum Computing.
PRC acknowledges support from the Swiss National Science Foundation under the Sinergia Grant CRSII5\_213497 (GW-Learn).
RS and LG acknowledge support from the PRIN2022 MUR project 2022CB3PJ3 - First Light And Galaxy aSsembly (FLAGS) funded by the European Union - Next Generation EU.
AL, AT, and RV acknowledge support by the PRIN MUR ``2022935STW'' funded by European Union-Next Generation EU, Missione 4 Componente 2, CUP C53D23000950006.
AT acknowledges financial support from the Bando Ricerca Fondamentale INAF 2023 Mini-grant ``Cosmic Archaeology with the first black hole seeds".
LM acknowledges support from the Swiss National Science Foundation under the Grant 200020\_207406.
RV acknowledges support from the Bando Ricerca Fondamentale INAF 2023, Theory Grant ``Theoretical models for Black Holes Archaeology''.
TZ and PRC thank Jillian M. Bellovary for discussions on the \textsc{Gasoline} code.
\end{acknowledgements}

\bibliographystyle{aa}
\bibliography{bibliography}

\clearpage

\begin{appendix}

\section{Code improvements and bug fixes}
\label{sec:code_improvements}

Initial tests revealed numerical issues with the earlier BH-evolution prescription. The previous version of the code, used in \citetalias{Sassano_et_al_2023}, was not optimised to investigate regimes as extreme as those considered here, both in terms of accretion rate (SE prescription) and mass resolution ($25~\msun$). Consequently, several modifications to the code were necessary to correctly model the systems under investigation and isolate the impact of numerical effects.

We detail below the most significant modifications made to the code, aimed primarily at resolving two issues:

\begin{enumerate}
    \item guaranteeing that a sufficient number of candidate gas particles are available for accretion;
    \item avoid cases wherein all particles in the BH vicinity are accreted in a single time step, ensuring that a minimum number of neighbouring particles is retained to allow proper feedback energy release.
\end{enumerate}

Accretion onto the BH is carried out using the nearest gas particles in the SPH scheme. Specifically, a fixed number of neighbouring particles, \( N_{\rm smooth} \), is selected to compute the local gas density \( \rho_{\rm gas} \) around the BH, which is then used to evaluate the accretion rate $\dot{M}_{\rm BHL}$ via Eq.~\eqref{eq:BHL}. The same \( N_{\rm smooth} \) particles also serve as the reservoir for both accretion and the subsequent injection of feedback energy, \( \epsilon_{\rm c } \epsilon_{\rm r} \dot{M}_{\rm BHL} \, {\rm d}t_{\rm eff} \, c^2 \), where ${\rm d}t_{\rm eff}$ is the effective integration time of the BH. 

\subsection{First issue: Identification of enough gas particles}

Due to the intrinsic nature of the SPH scheme, the mechanism just described can lead to a situation wherein a significant fraction of the $N_{\rm smooth}$ particles closest to the BH are, in fact, arbitrarily far from the BH itself -- i.e. at distances causally disconnected from the compact object, relative to the accretion timescale ${\rm d}t_{\rm eff}$. To prevent this, the previous version of the code implemented an if-clause to exclude gas particles that were too distant \citep[][]{Bellovary_et_al_2013}. Specifically, this prescription required two conditions to be satisfied to select a candidate gas particle:

\begin{itemize}
    \item The gas particle had to be part of the $N_{\rm smooth}$ nearest neighbours of the BH.
    \item The BH itself had to be among the $N_{\rm smooth}$ nearest neighbours of the gas particle.
\end{itemize}
This resulted in accreting only particles located within a sphere centred on the BH, with a radius smaller than the particle smoothing length $h$.\footnote{The smoothing length is defined in \textsc{Gasoline2} as half of the distance to the furthest neighbour and in our runs cannot be smaller than 10\% of the gravitational softening length.}

This approach led to severe numerical bottlenecks in two cases: in high-density environments, such as the gas-rich central region of a proto-galactic disc, and at very high resolution, when the typical smoothing length becomes much smaller than the typical distance between the BH and the nearest gas particles. This drastically suppressed accretion in such environments.

In the new version of the code, we initially replaced this condition with a much shallower criterion, whereby particles are marked as candidates for accretion if their distance from the BH is smaller than the maximum between their smoothing length and twice their (fixed) gravitational softening length $\epsilon$, i.e.
\begin{equation}
   d_i \leq \max\left(h_i, 2\epsilon\right),
\end{equation}
where $d_i$ is the distance between the $i$-th gas particle and the BH and $h_{i}$ is its smoothing length.

However, even this shallower criterion has been eventually removed, allowing all neighbouring particles to be accreted when needed, to ensure consistency between the integrated accretion rate and the BH mass growth.
We performed several tests to evaluate the maximum distance at which gas particles are selected for accretion, and found that, although in the vast majority of cases gas particles are located within a fraction of a parsec from the BH, in some simulations -- particularly those with strong feedback -- they can occasionally be accreted from a maximum distance of $\sim 2$~pc.
These events are rare, and the overall behaviour confirms the consistency of our approach.

\subsection{Second issue: Ensuring enough particles for feedback energy release}

Following the previous modification, in high-density environments, a sufficient amount of gas is selected to satisfy the accretion condition
\begin{equation}
    \Delta M_{\rm BH} = \dot{M}_{\rm BHL} {\rm d}t_{\rm eff} (1 - \epsilon_{\rm r}).
\end{equation}

At this point, a new problem can arise when high accretion rates would require the BH to absorb almost or all the entire combined mass of its neighbouring gas particles. This would leave no particles available to receive the feedback energy.

In the new version of the code, an upper limit is introduced on the accretion rate. Specifically, $\dot{M}_{\rm acc}$ is set to the minimum between ($i$) the BHL rate $\dot{M}_{\rm BHL}$, computed using the standard SPH algorithm applied to the $N_{\rm smooth}$ nearest neighbours of the BH, and ($ii$) a fixed fraction $\xi$ of the total gas mass available amongst the $N_{\rm smooth}$ particles over the timestep ${\rm d}t_{\rm eff}$:
\begin{equation}
   \dot{M}_{\rm acc} = \min\left(\dot{M}_{\rm BHL}\,,\, \xi\frac{ \sum^{N_{\rm smooth}}_{i} m_i}{{\rm d}t_{\rm eff}}\right),
\end{equation}
where $m_i$ is the mass of the $i$-th gas particle, and $\xi = 0.9$ in this work.\footnote{Here, $\xi = 0.9$ was chosen as the largest value that keeps corrections to the accretion rate negligible, while still ensuring that feedback is always launched in all of our simulations.}

\section{Initial conditions}

\begin{figure}
    \centering
    \includegraphics[width=1\linewidth]{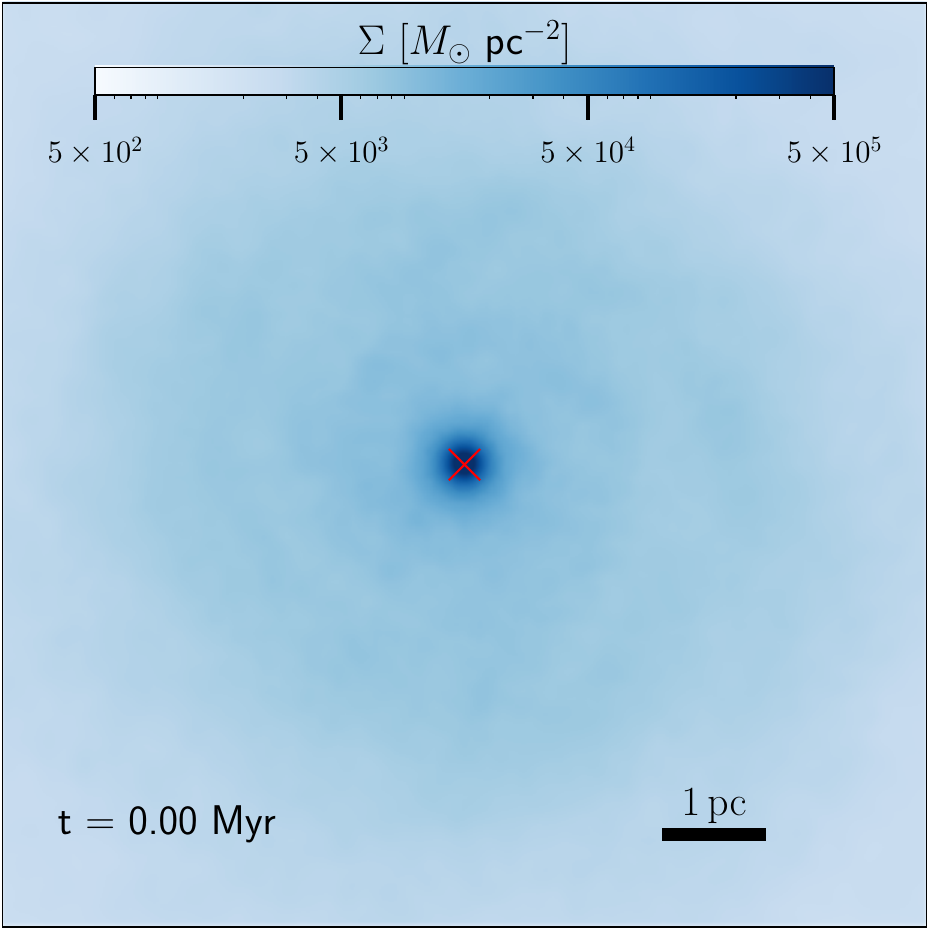}
    \caption{Surface density map ($\Sigma$) of the Mestel disc used as initial conditions in our simulations. The seeded BH is located at the centre and marked by a red cross.}
    \label{fig:map_ic}
\end{figure}

In Fig.~\ref{fig:map_ic}, we present the surface density map of the initial conditions adopted in the case $M_{\rm BH}(t_{\rm i})=5 \times 10^3$ as an example of the full set of simulations carried out in this study.
The region where the BH is seeded clearly shows a high central density of $2.4 \times 10^5~\msun\,\mathrm{pc}^{-3}$ within the central pc, which supports the early growth of the BH seed.

\section{Black hole accretion and migration}

\begin{figure*}
    \centering
    \includegraphics[width=1\linewidth]{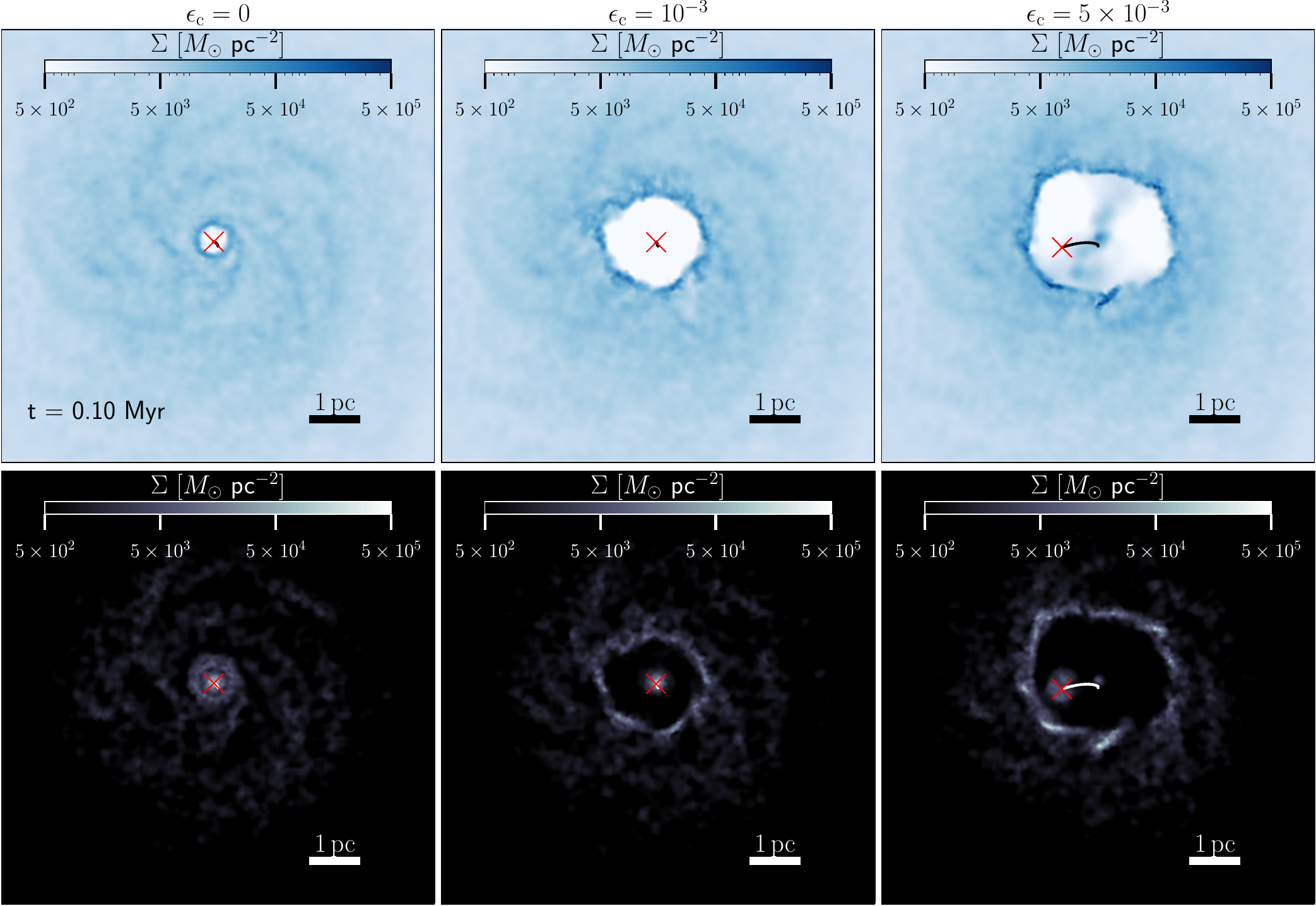}
    \caption{Surface density maps ($\Sigma$) of the gas (top row) and stellar (bottom row) components of a snapshot at $t = 0.1$~Myr of the same simulations shown in Fig.~\ref{fig:6maps_150}, i.e. with $M_{\rm BH}(t_{\rm i})=3 \times 10^5~\msun$ for the three feedback regimes of interest. The BH orbit is traced in black (top row) and white (bottom row), with the current BH position marked by a red cross.}
    \label{fig:6maps_10}
\end{figure*}

In Fig.~\ref{fig:6maps_10}, we show an earlier snapshot compared to Fig.~\ref{fig:6maps_150}, for the three values of $\epsilon_{\rm c}$ considered in our analysis.
At this stage, corresponding to $10^{-1}$~Myr from the beginning of the run, the first stars are starting to cumulate, and the gas density has begun to decline, although it has not yet been fully depleted.
The effects of feedback on the evacuation of the central region are clearly visible in the cases intermediate- and high-feedback.
Although the evacuated region appears more extended in the latter case, a faint gas bridge can still be seen feeding the BH.
In the no-feedback case, a central depleted region is still visible, entirely due to BH accretion.

In the rightmost panel, it is clear that the BH has received a kick which displaced it already by approximately 1~pc from the disc centre, in agreement with the discussion in Sec.~\ref{subsec:stalled_growth}.

Finally, where the remaining gas becomes denser, some spiral arms emerge in both the gas and stellar components.

Feedback has led to the accumulation of gas in a ring-like structure at a distance that depends on the strength of $\epsilon_{\rm c}$.
This accumulation has triggered localised SF, resulting in the emergence of stellar clumps that contribute to the scattering of the migrating BH as previously discussed.

\begin{figure}
    \centering
    \includegraphics[width=1\linewidth]{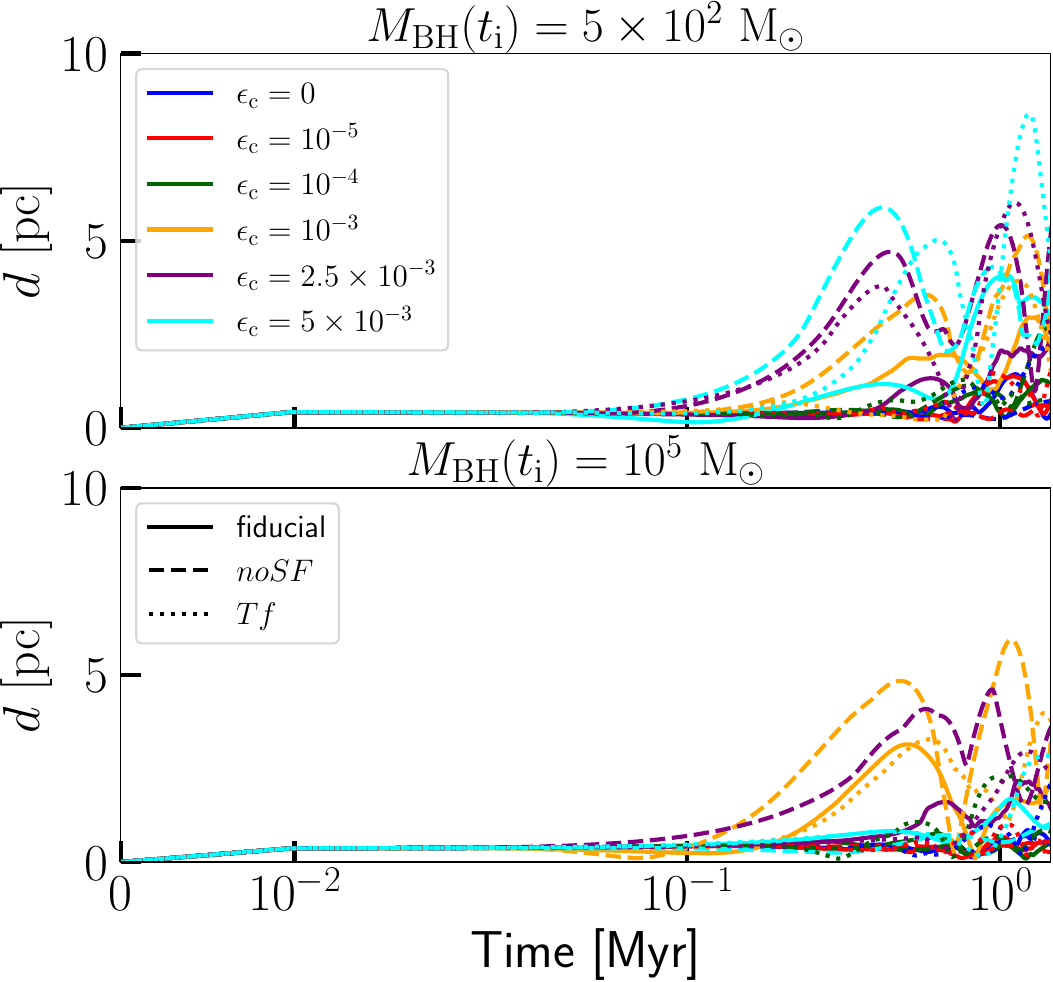}
    \caption{Evolution of the distance from the galactic centre for the suite of runs with initial BH masses $M_{\rm BH}(t_{\rm i}) = 5 \times 10^2~\msun$ (top panel) and $M_{\rm BH}(t_{\rm i}) = 10^5~\msun$ (bottom panel). Solid lines refer to our fiducial simulations, dashed lines indicate the noSF runs (without SF), and dotted lines correspond to the runs with both SF suppressed and an increased temperature floor (Tf).}
    \label{fig:dist_vs_noSF}
\end{figure}

The effect of SF is to slightly reduce the migration of the BH as discussed in Sec.~\ref{subsec:stalled_growth}.
In Fig.~\ref{fig:dist_vs_noSF}, we show the BH migration in terms of distance from the centre of the disc for the same cases analysed in Fig.~\ref{fig:massgain_vs_noSF}.
Aside from the orbital oscillations, it is evident that the final displacement of the BH increases significantly with feedback intensity. However, the presence of stars (and, to a lesser extent, gas clumps) has the opposite effect: it contributes to binding the BH more tightly, thereby reducing the amplitude of its motion.

\end{appendix}

\end{document}